\documentclass{emulateapj}
\usepackage{subfigure}
\usepackage{graphics}
\usepackage{natbib}
\usepackage{hyperref}
\newcommand{\teff}{T_{\mathrm{eff}}}
\newcommand{\feh}{\mathrm{[Fe/H]}}

\newcommand{\logg}{\log g}

\newcommand{\los}{\mathrm{los}}

\usepackage{amsmath}

\newcommand{\deriv}{\mathrm{d}}
\newcommand{\gradientvmean}{$87.5_{-0.4}^{+0.4}$}

\newcommand{\gradientvdisp}{$2.7_{-0.3}^{+0.3}$}

\newcommand{\gradientfehmean}{$-1.98_{-0.1}^{+0.1}$}

\newcommand{\gradientfehdisp}{$0.22_{-0.03}^{+0.04}$}
\newcommand{\gradientfehgrad}{$0.001_{-0.005}^{+0.005}$}

\newcommand{\gradientvmeanexpanded}{$87.5_{-0.4(-0.8)}^{+0.4(+0.8)}$}
\newcommand{\gradientvvarexpanded}{$0.9_{-0.1(-0.2)}^{+0.1(+0.2)}$}
\newcommand{\gradientvdispexpanded}{$2.7_{-0.3(-0.5)}^{+0.3(+0.7)}$}
\newcommand{\gradientvgradexpanded}{$0.03_{-0.02(-0.03)}^{+0.03(+0.05)}$}
\newcommand{\gradientvthetaexpanded}{$-103.0_{-51.0(-72.0)}^{+246.0(+279.0)}$}
\newcommand{\gradientfehmeanexpanded}{$-1.98_{-0.1(-0.2)}^{+0.1(+0.21)}$}
\newcommand{\gradientzvarexpanded}{$-1.3_{-0.14(-0.27)}^{+0.15(+0.28)}$}
\newcommand{\gradientfehdispexpanded}{$0.22_{-0.03(-0.06)}^{+0.04(+0.09)}$}
\newcommand{\gradientfehgradexpanded}{$0.001_{-0.005(-0.01)}^{+0.005(+0.01)}$}
\newcommand{\gradientbigsigmaexpanded}{$-0.57_{-0.05(-0.1)}^{+0.05(+0.09)}$}
\newcommand{\gradientlogrsexpanded}{$1.47_{-0.03(-0.07)}^{+0.03(+0.07)}$}
\newcommand{\gradientrsexpanded}{$29.24_{-2.19(-4.19)}^{+2.4(+4.95)}$}
\newcommand{\gradientbigsigmabexpanded}{$-0.64_{-0.0(-0.01)}^{+0.0(+0.01)}$}
\newcommand{\gradientbigsigmagradexpanded}{$0.0005_{-0.0001(-0.0002)}^{+0.0001(+0.0002)}$}
\newcommand{\gradientgradthetaexpanded}{$-87.4_{-8.46(-17.63)}^{+8.76(+17.72)}$}

\newcommand{\gradientvgradupperlim}{$0.06$}

\newcommand{\pmmualpha}{$-19.0_{-17.0}^{+17.0}$}
\newcommand{\pmmudelta}{$-14.0_{-19.0}^{+19.0}$}

\newcommand{\nrgb}{$10680$}

\newcommand{\nspec}{$ 390$}
\newcommand{\minsigv}{$0.3$}
\newcommand{\maxsigv}{$3.2$}
\newcommand{\medsigv}{$0.7$}
\newcommand{\minsigz}{$ 0.04$}
\newcommand{\maxsigz}{$ 0.92$}
\newcommand{\medsigz}{$ 0.13$}
\newcommand{\minsiglogg}{$0.1$}
\newcommand{\maxsiglogg}{$1.1$}
\newcommand{\medsiglogg}{$0.3$}
\newcommand{\minsigteff}{$  29$}
\newcommand{\maxsigteff}{$1085$}
\newcommand{\medsigteff}{$  93$}
\newcommand{\nmem}{$62.2_{- 0.6}^{+ 0.9}$}
\newcommand{\nmemfifty}{$  62$}

\newcommand{\ssppdeltav}{$ 0.98$}
\newcommand{\ssppdeltateff}{$ -397$}
\newcommand{\ssppdeltalogg}{$-0.90$}
\newcommand{\ssppdeltafeh}{$-0.58$}

\newcommand{\ssppsigmav}{$ 0.25$}
\newcommand{\ssppsigmateff}{$   11$}
\newcommand{\ssppsigmalogg}{$ 0.02$}
\newcommand{\ssppsigmafeh}{$ 0.01$}

\newcommand{\cramrhalf}{$4.4_{-0.9}^{+1.2}\times 10^6$}
\newcommand{\cravcirc}{$4.3_{-0.5}^{+0.5}$}
\newcommand{\cramwolf}{$7.0_{-1.5}^{+1.9}\times 10^6$}
\newcommand{\cravcircwolf}{$4.8_{-0.5}^{+0.6}$}
\newcommand{\cramlratioscale}{$10_{-2}^{+3}$}
\newcommand{\cramlratio}{$53_{-11}^{+15}$}
\newcommand{\cramlwolf}{$85_{-18}^{+25}$}

\newcommand{\cragalacticl}{$282.908$}
\newcommand{\cragalacticb}{$42.028$}
\newcommand{\cravgrf}{$-74.0_{-1.6}^{+1.5}$}

\def\kms{~km~s$^{-1}$\ }

\def\arcsec{\char'175 }

\def\hub{\ifmmode H_\circ\else H$_\circ$\fi}
\def\lsim{\mathop{\hbox{${\lower3.8pt\hbox{$<$}}\atop{\raise0.2pt\hbox{$\sim$}}$}}}
\def\gsim{\mathop{\hbox{${\lower3.8pt\hbox{$>$}}\atop{\raise0.2pt\hbox{$\sim$}}$}}}

\shorttitle{Kinematics of the Crater 2 Dwarf Galaxy}
\shortauthors{Caldwell et al}

\begin{document}

\title{Crater 2: An Extremely Cold Dark Matter Halo\footnote{Observations reported here were obtained at the MMT Observatory, a joint facility of the Smithsonian Institution and the University of Arizona.}}

\author{Nelson Caldwell\altaffilmark{1}, Matthew G. Walker\altaffilmark{2}, Mario Mateo\altaffilmark{3}, Edward W. Olszewski\altaffilmark{4}, Sergey Koposov\altaffilmark{5}, Vasily Belokurov\altaffilmark{5}, Gabriel Torrealba\altaffilmark{5}, Alex Geringer-Sameth\altaffilmark{2} and Christian I. Johnson\altaffilmark{1}}
\email{caldwell@cfa.harvard.edu}

\altaffiltext{1}{Harvard-Smithsonian Center for Astrophysics, 60 Garden Street, Cambridge, MA 02138, USA}
\altaffiltext{2}{McWilliams Center for Cosmology, Department of Physics, Carnegie Mellon University, 5000 Forbes Ave., Pittsburgh, PA 15213, United States}
\altaffiltext{3}{Department of Astronomy, University of Michigan, 311 West Hall, 1085 S. University Ave., Ann Arbor, MI 48109}
\altaffiltext{4}{Steward Observatory, The University of Arizona, 933 N. Cherry Ave., Tucson, AZ 85721}
\altaffiltext{5}{Institute of Astronomy, University of Cambridge, Madingley Road, Cambridge, CB3 0HA, United Kingdom}
\shorttitle{Crater 2: An Extremely Cold Dark Matter Halo} 

\begin{abstract}
  We present results from MMT/Hectochelle spectroscopy of \nspec\ red giant candidate stars along the line of sight to the recently-discovered Galactic satellite Crater 2.  Modelling the joint distribution of stellar positions, velocities and metallicities as a mixture of Crater 2 and Galactic foreground populations, we identify $\sim$ \nmemfifty\ members of Crater 2, for which we resolve line-of-sight velocity dispersion $\sigma_{v_{\rm los}}=$\gradientvdisp\ km s$^{-1}$ about mean velocity of $\langle v_{\los}\rangle$=\gradientvmean\ km s$^{-1}$ (solar rest frame).  We also resolve a metallicity dispersion $\sigma_{\mathrm{[Fe/H]}}$$=$\gradientfehdisp\ dex about a mean of $\langle$[Fe/H]$\rangle=$ \gradientfehmean\ dex that is $0.28\pm 0.14$ dex poorer than is estimated from photometry.  Despite Crater 2's relatively large size (projected halflight radius $R_{\rm h}\sim 1$ kpc) and intermediate luminosity ($M_V\sim -8$), its velocity dispersion is the coldest that has been resolved for any dwarf galaxy.  These properties make Crater 2 the most extreme low-density outlier in dynamical as well as structural scaling relations among the Milky Way's dwarf spheroidals.  Even so, under assumptions of dynamical equilibrium and negligible contamination by unresolved binary stars, the observed velocity distribution implies a gravitationally dominant dark matter halo, with dynamical mass \cramrhalf\ $M_{\odot}$ and mass-to-light ratio \cramlratio\ $M_{\odot}/L_{V,\odot}$ enclosed within a radius of $\sim 1$ kpc, where the equivalent circular velocity is \cravcirc\ km s$^{-1}$.
\end{abstract}


\section{Introduction}\label{intro}

The Milky Way's dwarf-galactic satellites represent an extremum of galaxy formation.  They include the smallest, least luminous, least chemically enriched and most dark-matter dominated galaxies known.  Their abundances as functions of size, luminosity, metallicity and dark matter density are sensitive to the nature of dark matter (e.g., the ability of dark matter particles to cluster on scales of $\sim$ tens of pc) as well as to how galaxy formation proceeds within low-mass dark matter halos \citep[e.g., ][]{pontzen14,dicintio14,read16}.  Empirical information has increased dramatically during the last decade, as deep imaging surveys have grown the number of known Galactic satellites from $\sim 10$ to several tens \citep[e.g., ][]{belokurov07,koposov15,des15,des15b} and spectroscopic campaigns have measured stellar kinematics and chemical abundances for tens to thousands of objects per system \citep[e.g., ][]{battaglia06,martin07,simon07,walker09a,kirby10}.  

The combination of structural and chemodynamical parameters from photometric and spectroscopic surveys, respectively, reveals empirical scaling relations that serve as targets for models of dwarf galaxy formation and evolution.  For example, the size/luminosity relation traced by dwarf spheroidals extends the one observed for disk galaxies but is offset from relations traced by central bulges and larger elliptical galaxies \citep{kormendy12}.  Moreover, the least luminous dwarf galaxies also exhibit the lowest metallicities, extending the galactic luminosity/metallicity relation over $\sim 5$ orders of magnitude in luminosity \citep{kirby13}.  Finally, correlations among internal line-of-sight velocity dispersion, size, and luminosity have been interpreted as evidence for similarity among the dark matter halos that dwarf galaxies inhabit \citep{mateo93,strigari08,walker09d}.  

While it is clear that dwarf galaxies' structural and chemodynamical properties are interconnected, physical interpretations are complicated by our ignorance regarding the extent to which these empirical correlations are driven by selection effects.  At a given distance and luminosity, smaller galaxies have higher surface densities and hence are detected more easily than are larger galaxies.   Careful quantification of sensitivities for various imaging surveys \citep[e.g., ][]{koposov08} leaves open the possibility that large numbers of large, low-luminosity galaxies exist undetected, perhaps modifying what are apparent size/luminosity correlations and/or other relations involving quantities that correlate with size and luminosity.  

In fact dwarf galaxies continue to be discovered down to the faintest surface brightnesses to which current imaging surveys are sensitive.  Most recently, \cite[T16 hereafter]{torrealba16} used data from the ATLAS survey \citep{shanks15} to discover Crater 2, a dwarf galaxy at distance $D\sim 117$ kpc from the Sun.  With absolute magnitude $M_V\sim -8.2$ and projected halflight radius $R_{\rm h}\sim 1.1$ kpc, Crater 2 exhibits the faintest surface brightness of any known dwarf galaxy, with $\mu_V\sim 31$ mag arcsec$^{-2}$.  Relative to other Galactic satellites of similar luminosity (or size), Crater 2 is exceptionally large (or underluminous).  

In order to gauge how this outlier status translates into chemodynamical scaling relations, and in order to study the dark matter content of Crater 2 in general, here we present results from initial MMT/Hectochelle spectroscopic observations of \nspec\ stars along the line of sight to the galaxy.  From the spectroscopic data, we estimate line-of-sight velocity as well as effective temperature, surface gravity and metallicity for each individual star.  We then use these measurements to characterise Crater 2's internal chemodynamics, to study its dark matter content, and to place Crater 2 in the context of the population of known Milky Way satellites.

\section{Observations and Data Reduction}\label{sec:obs}
We observed Crater 2 using the Hectochelle multi-object fiber spectrograph \citep{szentgyorgyi11} on the 6.5m MMT on Mt. Hopkins, Arizona.  Hectochelle's field of view subtends 1\degr\ in diameter and thus is well-matched to Crater 2's projected halflight radius of $R_{\rm h}\sim 31'$ (T16).  Hectochelle has been widely used for the study of MW dwarf galaxies \citep[e.g.,][]{mateo08,belokurov09,walker09c}, and achieves velocity precision better than 1 \kms \citep{walker15}. 

\subsection{Target Selection}
\begin{figure*}
  \centering
  \begin{tabular}{@{}ll@{}}
    \includegraphics[width=2.5in, trim=0in 2.5in 2.75in 0in,clip]{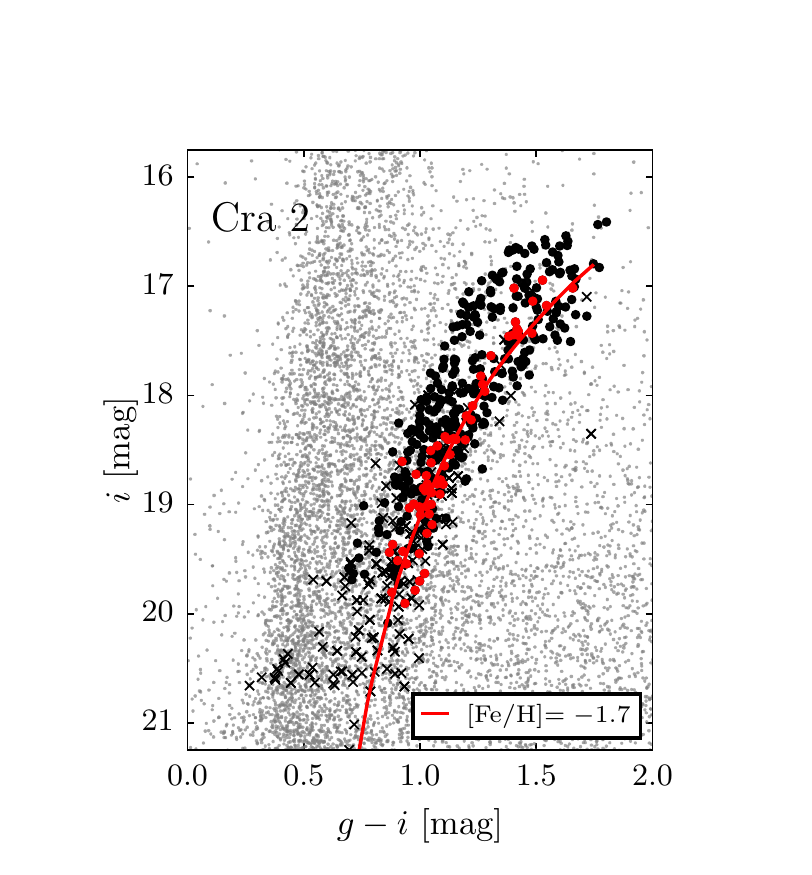}&\includegraphics[width=4.2in,trim=0in 2.5in 0in 0.5in,clip]{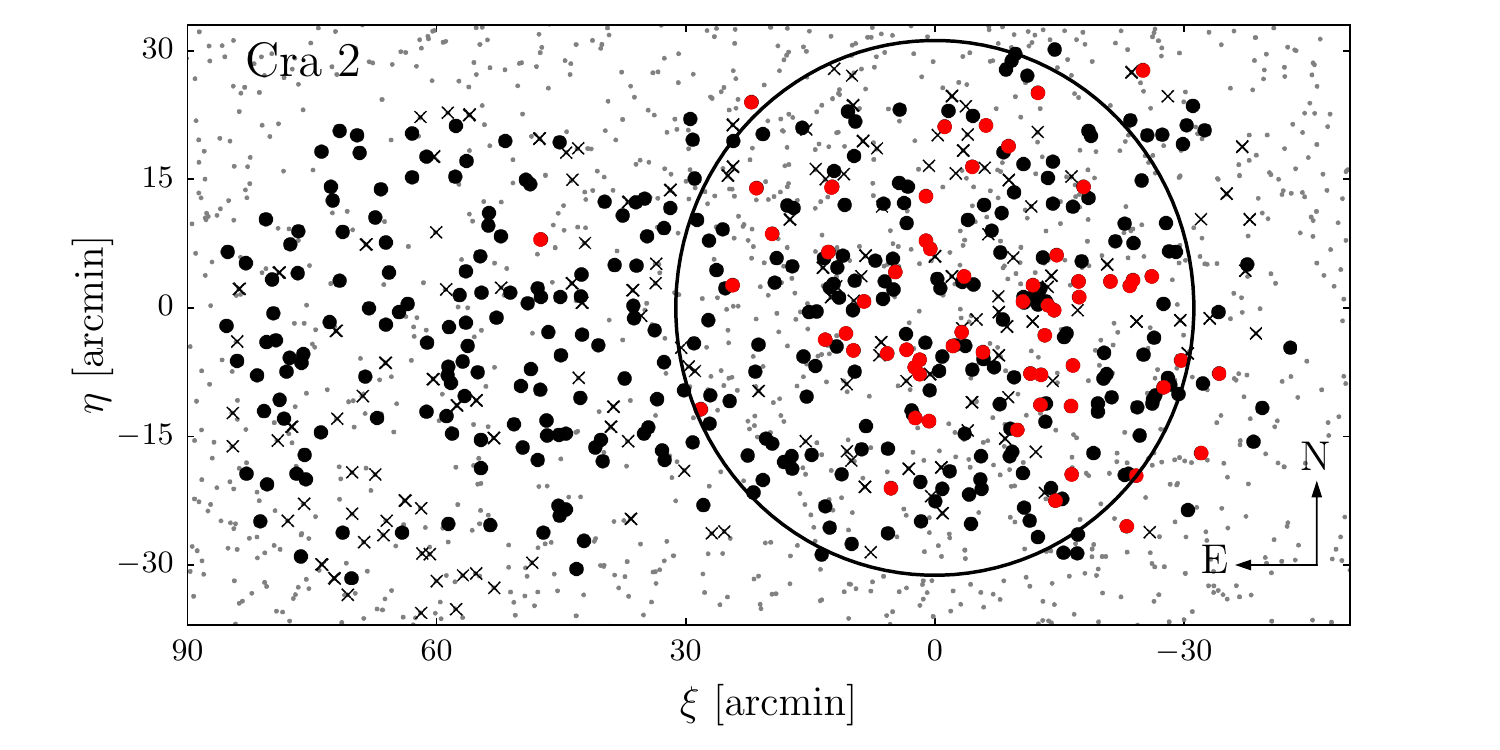}
  \end{tabular}
  \caption{\textit{Left:} Color-magnitude diagram for stars within $R\leq 0.5^{\circ}$
    of the center of Crater 2 \citep[`T16']{torrealba16}.  The red line shows the
    Dartmouth isochrone \citep{dotter08} for age$=12$ Gyr, [Fe/H]$=-1.7$,
    [$\alpha/$Fe]$=+0.4$, and distance modulus $m-M=20.3$.    
    \textit{Right:} Positions for stars photometrically selected as red giant branch
    candidates.  The black circle marks the 2D halflight radius
    measured by T16.  In both
    panels, filled red/black circles represent probable members/nonmembers in
    our Hectochelle spectroscopic catalog (see Section \ref{subsec:membership}); black `x's mark stars with spectra that did not pass quality-control criteria.
    \label{fig:cra2_cmd}}
\end{figure*}

In order to identify stars for spectroscopic targeting, we selected red giant branch (RGB) candidates from the photometric catalog provided by the VLT/ATLAS survey \citep{shanks15}.  The left panel of Figure \ref{fig:cra2_cmd} shows the color-magnitude diagram for point sources near the center of Crater 2.  The overplotted isochrone \citep{dotter08} corresponds to age=12 Gyr,  [Fe/H]=$-1.7$ and distance modulus $m-M=20.3$, closely following the ridge line identified by T16.  In addition to bona fide red giants at the distance of Crater 2, this isochrone likely coincides with less-luminous but nearer dwarf stars in the Milky Way foreground.  In order to avoid biasing our spectroscopic observations in metallicity, we selected RGB candidates spanning a range of $\sim 0.5$ magnitudes in color around the isochrone.  The right panel of Figure \ref{fig:cra2_cmd} shows positions of these RGB candidates on the sky.  In both panels, larger markers identify the RGB candidates for which we obtained spectra of sufficient quality to present here.  Having obtained a catalog of RGB candidates, we allocated fibers to individual stars, giving lower priority to fainter targets.  

%Hectochelle is a single order multi-object spectrograph, where an order is about 150\AA \ and is selected by a filter.  For our Crater 2 observations, we use the order spanning approximately 5150 - 5300 \AA, containing the Mgb triplet, as it is optimal for RGB stars.  For hot stars, however, this region contains virtually no absorption features.  For this reason, even though T16 detect BHB populations in Crater 2, we did not target BHBs for this program.  Otherwise

%\begin{figure}[ht]
%\includegraphics[width=3.3in]{cmd_input.pdf} 
%\caption{CMD of input catalog. 
%\label{input}}
%\end{figure}

\subsection{Observations and Data Reduction}\label{subsec:obs}

On the nights of 27 April, 30 April and 11 May 2016, we used Hectochelle's `RV31' filter to observe the spectral range $5150 - 5300$ \AA, containing the prominent Mg~b triplet, for three fields in Crater 2.  Two fields provided different sets of targets in the center of Crater 2, and the third field was offset from the center by 1.5\degr to the east (see right panel of Figure \ref{fig:cra2_cmd}).  Exposure times ranged between $2-3$ hours per field.  Table \ref{fields} provides details of the exposures, which provided spectra for 610 unique science targets.   In each field, $\sim 40$ fibers observe regions of blank sky in order to facilitate sky subtraction.  

In addition to science exposures, we also acquired exposures of ThAr arc lamps, taken immediately before and after science exposures, for the purpose of wavelength calibration.  At the beginning of each night, we also obtained twilight sky exposures in order to correct for fiber throughput variations and calibrate our measurements of stellar-atmospheric parameters.  For all exposures, the Hectochelle CCDs were binned by 3 pixels in the spectral direction and by 2 in the spatial direction. The effective resolution is $R \approx 32,000$.  All data frames were processed using the TDC/Hectochelle pipeline, which provides 1D, throughput-corrected, wavelength-calibrated spectra, as well as their corresponding variance spectra, as described in detail by \cite{caldwell09}.  

\begin{deluxetable*}{lllllll}
%\rotate
%\tablenum{1}
\tablecolumns{7}
\tablewidth{0pc}
%\tablewidth{4truein}
\tablecaption{Fields observed \label{fields}} 
\tablehead{\colhead{Field \#} &\colhead{Date} &\colhead{Field Center} &\colhead{Order} & \colhead{Exptime}  & \colhead{\# targets}  &\colhead{\# good measurements} \\
\colhead{} & \colhead{} & \colhead{} & \colhead{} & \colhead{sec} & \colhead{} & \colhead{} } 
\startdata
%1 & 2016/01/28 & 11:49:05.46 ~~ $-$18:28:24.71 & Ca~T & 9817 & 164\\
1 & 2016/04/27 & 11:48:16.20  ~~   $-$18:24:14.53 & Mg~b & 10800 & 204&160\\
2 & 2016/04/30 &  11:50:02.56  ~~   $-$18:26:26.78 & Mg~b & 8100 & 202&106\\
3 & 2016/05/11 & 11:53:22.09 ~~    $-$18:30:43.74 & Mg~b & 8100 & 204&124\\
totals & & & & &610&390\\
\enddata
\end{deluxetable*}

Given the processed spectra for a given exposure, we follow the procedure of \citet{koposov11} to estimate the mean sky spectrum and to subtract it from each science spectrum.  All processed and sky-subtracted spectra for this program are publicly available and included in the online database associated with this article.

We analyze each sky-subtracted spectrum following the procedure of \citet{walker15}.  Briefly, we fit a smoothed library of synthetic template spectra, obtaining Bayesian inferences for the line-of-sight velocity ($v_{\rm los}$) as well as stellar-atmospheric parameters including effective temperature ($T_{\rm eff}$), surface gravity ($\log g$) and metallicity ([Fe/H]).  As in previous work, we use the synthetic library originally generated for the SEGUE Stellar Parameter Pipeline \citep[][`SSPP' hereafter]{lee08a,lee08b}, which contains continuum-normalized spectra computed over a regular grid in $T_{\rm eff}$, $\log$ g and [Fe/H] and assumes a piecewise-linear relation between [$\alpha$/Fe] and [Fe/H].  

We perform the fits using the nested-sampling algorithm MultiNest\footnote{available at ccpforge.cse.rl.ac.uk/gf/project/multinest} \citep{feroz08,feroz09}, which returns random samples from the posterior probability distribution function.  From these samples, we record the mean, variance, skewness and kurtosis that summarize the 1D posterior PDF for each free parameter.  Following \citet{walker15}, we enforce quality-control criteria by discarding observations for which the PDF for $v_{\rm los}$ is non-Gaussian, as quantified by skewness $S$ and kurtosis $K$.  That is, we retain only those observations for which $|S|\leq 1$ and $|K-3|\leq 1$.  

Finally, we adjust all means and variances, which represent our estimate and errorbar, according to results from our $\sim 500$ twilight spectra, from which we gauge zero-point offsets (and corresponding uncertainties) with respect to solar values.  The offsets that we obtain from the twilights observed during our Crater 2 observations are similar to those obtained during the Draco observations reported by \citealt{walker15}.  Specifically, we subtract the following values from our raw estimates of velocity, temperature, gravity and metallicity: $\Delta v_{\los}$$=$\ssppdeltav\ km s$^{-1}$, $\Delta\teff$$=$\ssppdeltateff\ K, $\Delta\logg$$=$\ssppdeltalogg\ dex, $\Delta\feh$$=$\ssppdeltafeh\ dex.  To the raw variances we add the (squares of) the following standard deviations obtained from the twilight spectra: $\sigma_{v_{\los}}$$=$\ssppsigmav\ km s$^{-1}$, $\sigma_{\teff}$$=$\ssppsigmateff\ K, $\sigma_{\logg}$$=$\ssppsigmalogg\ dex, $\sigma_{\feh}$$=$\ssppsigmafeh\ dex. 

\begin{figure}[ht]
\includegraphics[width=3.25in, trim=0in 0.5in 4.3in 3.in,clip]{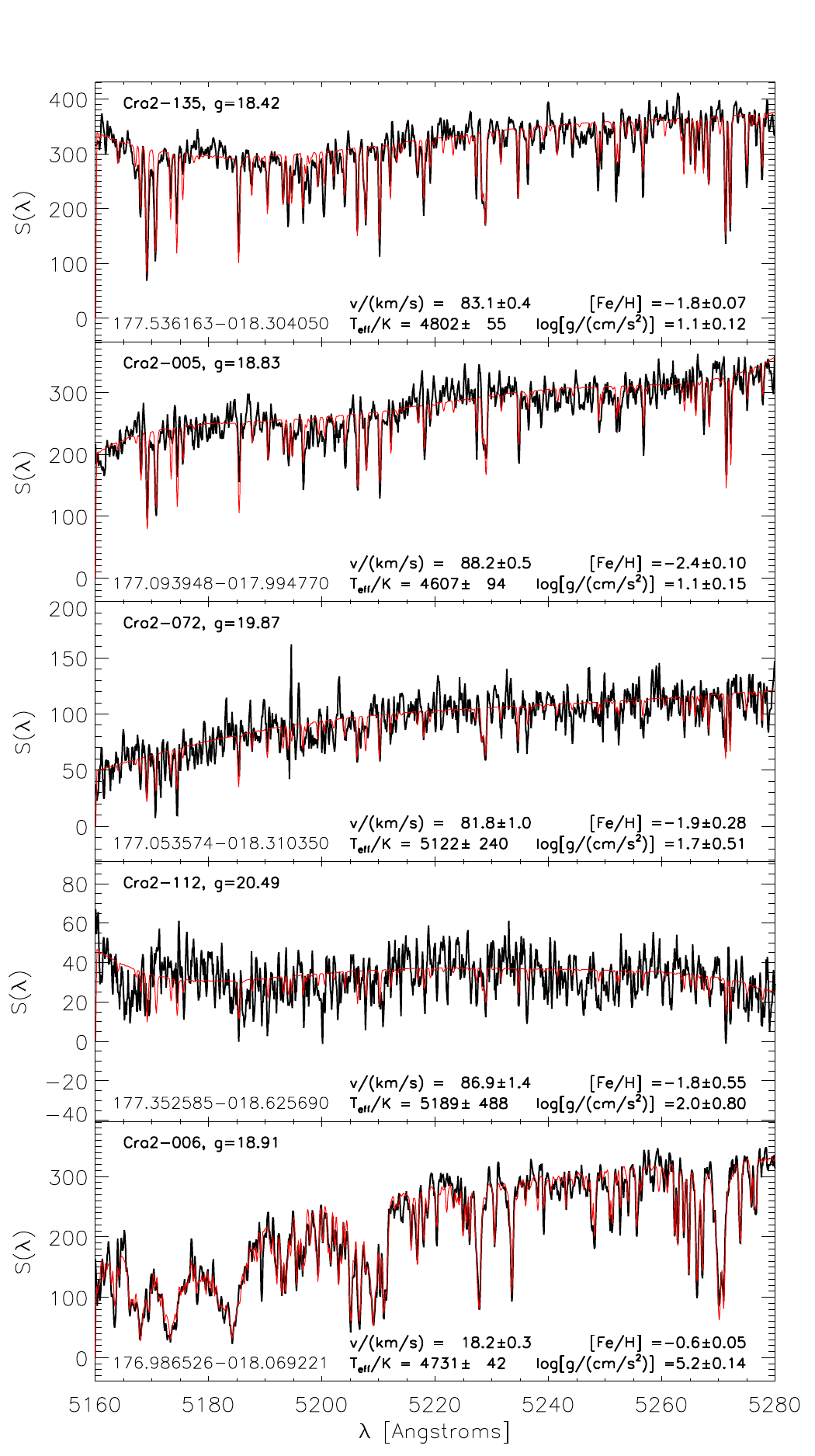} 
\caption{Examples of sky-subtracted MMT/Hectochelle spectra (black) for stars observed along the line of sight to Crater 2, with best-fitting models overplotted (red).  Spectra in the top four panels correspond to likely red-giants within Crater 2; the spectrum in the bottom panel is a metal-rich dwarf star in the Milky Way foreground.  Text in each panel indicates the star ID (see Table \ref{tab:cra2_table1}), $g$ magnitude from the ATLAS catalog, equatorial coordinates and our estimates of spectroscopic quantities.  
\label{fig:cra2_spectra}}
\end{figure}

\section{Estimates of Stellar Parameters}
\label{sec:data}

For each of the \nspec\ observations that satisfy our quality-control criteria, Table \ref{tab:cra2_table1} lists stellar position, $g$ and $i$ magnitudes from the ATLAS catalog, time of observation, signal-to-noise ratio, and our spectroscopic estimates of velocity, effective temperature, surface gravity and metallicity.  Our estimates of spectroscopic quantities have median (minimum, maximum) errors of $\sigma_{v_{\los}}$$=$\medsigv\ (\minsigv, \maxsigv) km s$^{-1}$, $\sigma_{\teff}$$=$\medsigteff\ (\minsigteff, \maxsigteff) K, $\sigma_{\logg}$$=$\medsiglogg\ (\minsiglogg, \maxsiglogg) dex and $\sigma_{\feh}$$=$\medsigz\ (\minsigz, \maxsigz) dex.  Along with the information in Table \ref{tab:cra2_table1}, all processed and sky-subtracted spectra, as well as random samplings of posterior PDFs generated by MultiNest, are included in the online database associated with this article.  
\begin{table*}
\scriptsize
\caption{Hectochelle Stellar Spectroscopy of Crater 2$^{a}$}
\begin{tabular}{@{}ccccccccccccccccccccccc@{}}
\hline
\\
ID&$\alpha_{2000}$&$\delta_{2000}$&$g$&$i$&HJD$^{b}$&S/N$^{c}$&$v_{\rm los}$&$T_{\rm eff}$&$\log_{10}[g/$(cm/s$^{2}$)]&$\mathrm{[Fe/H]}$\\

& [hh:mm:ss]&[$^{\circ}$:$\arcmin$:$\arcsec$]&[mag]&[mag]&[days]&&[km s$^{-1}$]$^{d}$&[K]&[dex]&[dex]\\
\hline
Cra2-002&11:49:50.39&-18:23:59.5&$ 18.85$&$ 17.44$&$  7505.64$&$  8.5$&$  86.2\pm       0.4$&$4598\pm  81$&$1.09\pm0.14$&$-1.86\pm0.09$\\%$1.000_{-0.000}^{+0.000}$
Cra2-003&11:46:59.26&-18:41:38.0&$ 18.87$&$ 17.45$&$  7505.64$&$  8.6$&$  94.2\pm       0.4$&$4547\pm  83$&$1.44\pm0.20$&$-1.98\pm0.09$\\%$0.999_{-0.001}^{+0.000}$
Cra2-004&11:49:22.24&-18:32:25.9&$ 18.84$&$ 17.46$&$  7505.64$&$  9.8$&$  89.5\pm       0.5$&$4779\pm  82$&$1.21\pm0.20$&$-2.00\pm0.10$\\%$1.000_{-0.000}^{+0.000}$
Cra2-005&11:48:22.55&-17:59:41.2&$ 18.83$&$ 17.41$&$  7505.64$&$ 10.2$&$  88.2\pm       0.5$&$4607\pm  94$&$1.12\pm0.15$&$-2.41\pm0.10$\\%$1.000_{-0.000}^{+0.000}$
Cra2-025&11:48:24.75&-18:22:08.5&$ 19.17$&$ 17.90$&$  7505.64$&$  9.0$&$  91.1\pm       0.5$&$4750\pm 100$&$1.28\pm0.24$&$-2.28\pm0.12$\\%$1.000_{-0.000}^{+0.000}$
Cra2-026&11:47:32.40&-18:44:16.4&$ 19.08$&$ 17.82$&$  7505.64$&$  8.6$&$  88.9\pm       0.6$&$4758\pm 104$&$1.12\pm0.17$&$-2.36\pm0.12$\\%$1.000_{-0.000}^{+0.000}$
Cra2-036&11:48:14.08&-18:25:02.8&$ 19.32$&$ 18.10$&$  7505.64$&$  7.7$&$  87.8\pm       0.5$&$4503\pm  73$&$1.17\pm0.18$&$-2.18\pm0.08$\\%$1.000_{-0.000}^{+0.000}$
Cra2-039&11:49:34.62&-18:20:33.7&$ 19.39$&$ 18.19$&$  7505.64$&$  6.5$&$  89.2\pm       0.5$&$4721\pm 110$&$1.30\pm0.24$&$-1.91\pm0.13$\\%$1.000_{-0.000}^{+0.000}$
Cra2-043&11:48:55.72&-18:08:18.4&$ 19.44$&$ 18.22$&$  7505.64$&$  7.4$&$  87.4\pm       0.6$&$4761\pm 111$&$1.42\pm0.27$&$-2.09\pm0.14$\\%$1.000_{-0.000}^{+0.000}$
Cra2-049&11:47:24.54&-18:21:05.2&$ 19.48$&$ 18.38$&$  7505.64$&$  7.0$&$  86.8\pm       0.5$&$4736\pm 120$&$1.44\pm0.29$&$-1.87\pm0.15$\\%$1.000_{-0.000}^{+0.000}$
Cra2-053&11:50:06.80&-18:10:44.4&$ 19.56$&$ 18.40$&$  7505.64$&$  5.0$&$  88.4\pm       0.8$&$4787\pm 155$&$1.39\pm0.32$&$-2.10\pm0.20$\\%$1.000_{-0.000}^{+0.000}$
Cra2-055&11:49:59.68&-18:27:44.9&$ 19.54$&$ 18.46$&$  7505.64$&$  4.6$&$  83.0\pm       0.7$&$4576\pm 121$&$1.29\pm0.26$&$-2.32\pm0.14$\\%$1.000_{-0.000}^{+0.000}$
Cra2-059&11:49:19.00&-18:11:45.5&$ 19.55$&$ 18.51$&$  7505.64$&$  6.1$&$  85.0\pm       0.8$&$4937\pm 164$&$1.31\pm0.31$&$-2.22\pm0.21$\\%$1.000_{-0.000}^{+0.000}$
Cra2-064&11:48:13.11&-18:47:10.0&$ 19.67$&$ 18.54$&$  7505.64$&$  5.2$&$  86.9\pm       0.7$&$4775\pm 148$&$1.56\pm0.36$&$-1.95\pm0.18$\\%$1.000_{-0.000}^{+0.000}$
Cra2-072&11:48:12.86&-18:18:37.3&$ 19.87$&$ 18.85$&$  7505.64$&$  4.8$&$  81.8\pm       1.0$&$5122\pm 240$&$1.66\pm0.51$&$-1.92\pm0.28$\\%$0.999_{-0.001}^{+0.000}$
Cra2-073&11:48:21.07&-18:36:04.0&$ 19.86$&$ 18.77$&$  7505.64$&$  4.7$&$  86.3\pm       0.7$&$4934\pm 148$&$1.73\pm0.39$&$-1.90\pm0.19$\\%$1.000_{-0.000}^{+0.000}$
Cra2-075&11:48:18.65&-18:27:54.7&$ 19.89$&$ 18.81$&$  7505.64$&$  2.8$&$  94.5\pm       1.1$&$5066\pm 334$&$2.01\pm0.60$&$-1.75\pm0.40$\\%$0.998_{-0.001}^{+0.001}$
Cra2-076&11:47:33.94&-18:21:29.7&$ 19.87$&$ 18.84$&$  7505.64$&$  4.3$&$  86.6\pm       0.8$&$4795\pm 164$&$1.69\pm0.48$&$-1.75\pm0.21$\\%$1.000_{-0.000}^{+0.000}$
Cra2-080&11:47:36.90&-18:50:09.0&$ 19.89$&$ 18.85$&$  7505.64$&$  3.5$&$  85.4\pm       0.9$&$4658\pm 164$&$1.33\pm0.30$&$-2.28\pm0.19$\\%$1.000_{-0.000}^{+0.000}$
Cra2-082&11:47:18.31&-18:33:57.7&$ 19.94$&$ 18.89$&$  7505.64$&$  3.8$&$  87.7\pm       0.8$&$4840\pm 167$&$1.36\pm0.33$&$-1.74\pm0.21$\\%$1.000_{-0.000}^{+0.000}$
\hline
\end{tabular}
\\
\raggedright
$^{a}$See electronic edition for complete data table.\\
$^{b}$ heliocentric Julian date minus $2.45\times 10^6$ days\\
$^{c}$median signal-to-noise ratio per pixel\\
$^{d}$line-of-sight velocity in the heliocentric rest frame\\
\label{tab:cra2_table1}
\end{table*}

\begin{figure}[ht]
\includegraphics[width=3.25in, trim=0in 0in 2.8in 0.6in,clip]{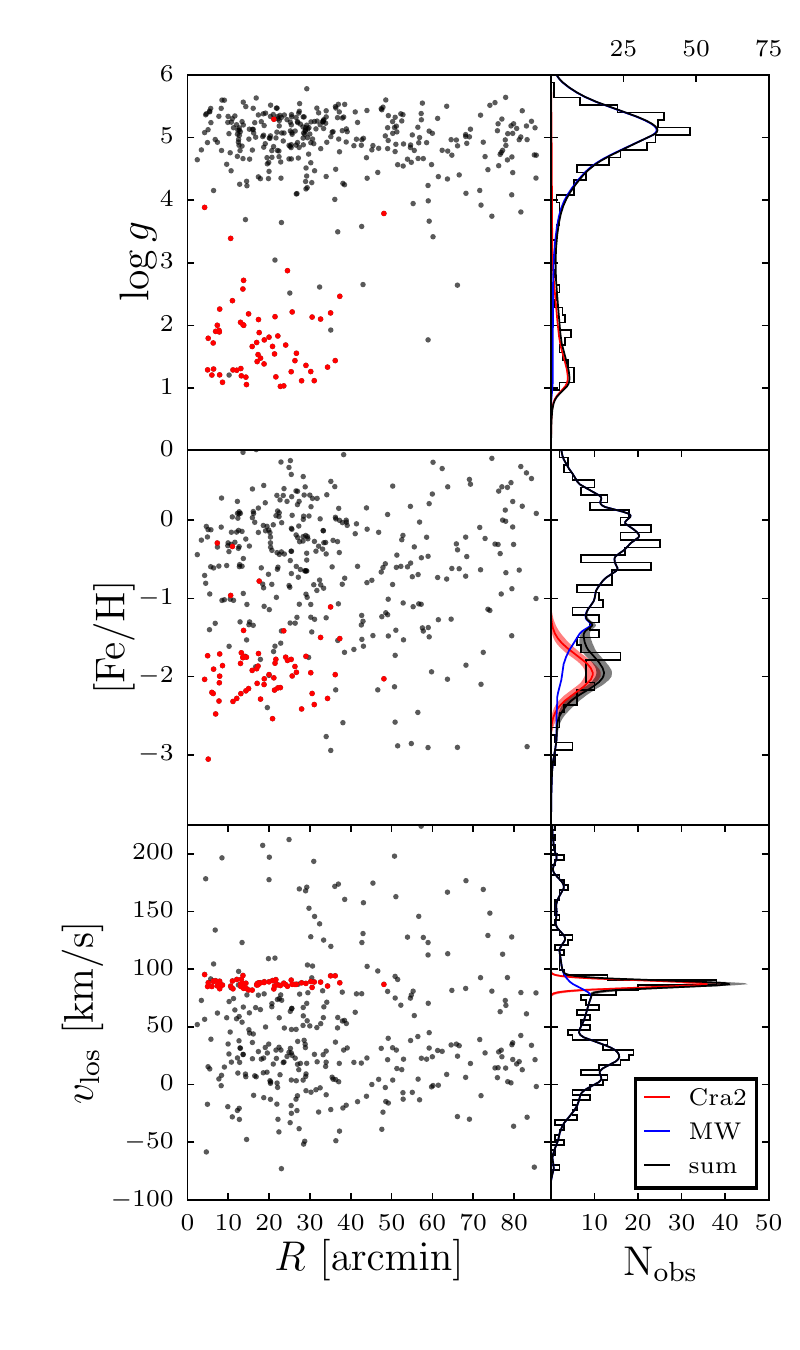} 
\caption{Spectroscopically-measured surface gravity (top), metallicity (middle) and velocity
  (bottom) vs angular separation from Crater 2's center.  Red (black)
  markers represent stars with posterior membership probability $\geq 50\%$
  ($<50\%$).  Histograms display 1D distributions, with posterior PDFs for the two populations (Crater 2 members and
  Milky Way foreground) of our mixture model overplotted.
\label{fig:cra2_scatter}}
\end{figure}

\section{Chemodynamics of Crater 2}
\label{sec:chemo}

The available data for Crater 2 consist of photometry from the ATLAS survey and our new Hectochelle spectroscopy.  From the photometric data we obtain a sample of projected positions, $\vec{R}$, for the $N_{\rm RGB}$$=$\nrgb\ RGB candidates, selected using the same isochrone mask employed by T16, that have projected positions within $R_{\rm max}$$=175$ arcmin of Crater 2's center at $\alpha_{\rm J2000}=177.310^{\circ}$, $\delta_{\rm J2000}=-18.413^{\circ}$.  For $N_{\rm spec}$$=$\nspec\ of these  RGB candidates, the spectroscopic data set provides measurements of LOS velocity $V$, metallicity $Z$, surface gravity $G$.  Given a model, $\Theta$, that specifies the projected stellar density, $\Sigma(\vec{R}|\Theta)$, of RGB candidates as well as the joint probability, $P(V,Z,G|\vec{R},\Theta)$, of spectroscopic quantities as a function of position, the two data sets have joint likelihood
\begin{eqnarray}
  \mathcal{L}\propto \exp\biggl [-\int_{\mathcal{R}}\deriv\vec{R}\,\Sigma(\vec{R}|\Theta)\biggr ]\prod_{i=1}^{N_{\rm RGB}}\Sigma(\vec{R}_i|\Theta)\nonumber\\
  \times \prod_{i=1}^{N_{\rm spec}}P(V_i,Z_i,G_i|\vec{R}_i,\Theta),
  \label{eq:like1}
\end{eqnarray}
where the constant of proportionality does not depend on the model.  The argument of the exponential factor is the (negative) expected number of RGB candidates counted over the 2D field $\mathcal{R}$; thus we model the count of RGBs inside any (infinitesimally) small area element within $\mathcal{R}$ as a Poisson random variable.  

For $\Theta$ we adopt a mixture model under which both data sets sample two stellar populations---Crater 2 members and Galactic foreground contamination---whose observables follow distinct chemodynamical distributions.  We assume that the 2D spatial distribution of the member population (indicated hereafter by subscript `1') follows the circularly symmetric \citet{plummer11} profile fit by \citet{torrealba16}:
\begin{equation}
  \Sigma_1(R|\Theta)=\Sigma_{0,1}\biggl [ 1+\frac{R^2}{R^2_{\rm h}}\biggr ]^{-2},
\end{equation}
where $\Sigma_{0,1}$ is the projected density at Crater 2's center and scale radius $R_{\rm h}$ is the 2D halflight radius---i.e., the radius of the circle enclosing half the member stars.  Also following \citet{torrealba16}, we assume the foreground population (indicated by subscript `2' hereafter) follows a 2D spatial distribution that varies linearly across the field.  We adopt the model
\begin{equation}
  \Sigma_2(R,\theta|\Theta)=\Sigma_{0,2}\bigl [1+k_{2}R\cos(\theta-\theta_{2})\bigr ],
\end{equation}
where $(R,\theta)$ are polar coordinates with origin at the center of Crater 2, $\Sigma_{0,2}$ is the projected density of foreground stars at the origin, and gradient $k_{2}$ and its direction $\theta_{2}$ are additional free parameters.  Given our mixture model, the conditional likelihood of spectroscopic quantities becomes a density-weighted sum of the probability distributions followed separately by each population:
\begin{eqnarray}  P(V,Z,G|\vec{R},\Theta)=\hspace{2.4in}\\
  \frac{\Sigma_1(R|\Theta)P_1(V,Z,G|\vec{R},\Theta)+\Sigma_2(R,\theta|\Theta)P_2(V,Z,G|\vec{R},\Theta)}{\Sigma_1(R|\Theta)+\Sigma_2(R,\theta|\Theta)}.\nonumber
\end{eqnarray}

We assume that, for both populations, distributions of velocities, metallicities and surface gravities are separable functions of position, such that $P_1(V,Z,G|\vec{R},\Theta)=P_1(V|\vec{R},\Theta)P_1(Z|\vec{R},\Theta)P_1(G|\vec{R},\Theta)$ (and similar for the foreground population).  We further assume that member velocities and metallicities follow Gaussian distributions, each with constant dispersion about means that vary smoothly with position.  In order to capture the effect of solid-body rotation and/or perspective-induced `rotation' due to projection of Crater 2's systemic proper motion \citep{feast61,kaplinghat08}, we allow a velocity gradient with magnitude $k_V\equiv \deriv V/\deriv R$ to point in a direction specified by position angle $\theta_V$.  We assume any metallicity gradient is azimuthally uniform and thus has only a magnitude, $k_Z\equiv \deriv Z/\deriv R$.  Thus, for a star at position $\vec{R}$, observations of velocity $V$ and metallicity $Z$, with respective measurement errors $\delta_V$ and $\delta_Z$, have probabilities
\begin{eqnarray}
  P_1(V|\vec{R},\Theta)=\mathcal{N}_V(\overline{V}_1-k_VR\cos(\theta_V-\theta_i),\sigma^2_V+\delta^2_V);\nonumber\\
  P_1(Z|\vec{R},\Theta)=\mathcal{N}_Z(\overline{Z}_1-k_ZR,\sigma^2_Z+\delta^2_Z),\hspace{0.82in}
\end{eqnarray}
where $\mathcal{N}_X(\overline{X},\sigma^2_X)\equiv (2\pi\sigma^2_X)^{-1/2}\exp[-\frac{1}{2}(X-\overline{X})^2\sigma^{-2}_X]$.  

Finally, we assume that distributions of the following are independent of position: surface gravities for both populations, and velocities and metallicities for the foreground population.  For the member population, this assumption is justified by Crater 2's extremely low surface brightness, which makes stellar encounters---and hence the mass segregation that would impart spatial dependence to the surface gravity distribution---negligible.  For the foreground, it is justified by the smallness of the field compared to scales over which the Galactic distributions vary.\footnote{These assumptions of position independence refer to the distributions intrinsic to the populations that we observe, but because of selection effects they do not necessarily hold for the samples that we actually acquire.  For example, our observations reach different limiting magnitudes (by up to $\sim 0.5$ mag) in each of the three observed fields, thereby imparting some spatial dependence to the observed distribution of $\logg$.  However, we confirm that we obtain nearly identical (except for slightly larger errorbars) results if we discard the spectroscopic data obtained for Fields 2 and 3, fitting only the data from the deepest (Field 1) observation.}  In order to prevent the number of free parameters from growing unwieldy, we estimate each of the position-independent distributions by smoothing the data with Gaussian kernels weighted by prior membership probabilities, $P_{\rm mem}$ (or non-membership probabilities, $P_{\rm non}=1-P_{\rm mem}$ as appropriate) obtained using the expectation-maximization algorithm as described by \citet{walker09b}.  For example, we estimate the distribution of surface gravities for Crater 2 members as
\begin{equation}
  \hat{P}_1(G)=\frac{\sum_{i=1}^{N_{\rm spec}}P_{\mathrm{mem}_i}\mathcal{N}_{G_i}(G,\delta^2_{G_i})}{\sum_{i=1}^{N_{\rm spec}}P_{\mathrm{mem}_i}},
\end{equation}
with measurement error $\delta_{G}$ serving as the smoothing bandwidth\footnote{In order to reduce noise in our estimate of the foreground velocity distribution, which is sparsely sampled, we smooth the data using a constant bandwidth of 5 km s$^{-1}$.  We confirm that our results are insensitive to this choice among other plausible bandwidths.}.

Again we use MultiNest to estimate parameters for the 12-dimensional model described above.  Table \ref{tab:cra2_gradient} lists these parameters as well as the adopted prior probability distributions, which are uniform over the indicated ranges and zero outside those ranges.  The  third column of Table \ref{tab:cra2_gradient} summarizes the marginalized, 1D posterior distributions returned by MultiNest, giving median-likelihood values as well as intervals that enclose the central 68\% and 95\% of posterior probability.  The bottom three rows of Table \ref{tab:cra2_gradient} give the corresponding constraints on posterior PDFs for quantities of interest (e.g., $\sigma_{v_{\los}}$, $\sigma_{\mathrm{[Fe/H]}}$, $R_{\rm h}$) that are functions of the free parameters ($\log_{10}[\sigma^2_V/(\mathrm{km^2s^{-2}})]$, $\log_{10}[\sigma_{\mathrm{[Fe/H]}}]$, $\log_{10}[R_{\rm h}/\mathrm{arcmin}]$, respectively)\footnote{Our results are insensitive to whether our uniform prior was applied to the linear quantities or to their logarithms.}.  

Plotted over the histograms of observed velocities, metallicities and surface gravities in Figure \ref{fig:cra2_scatter} are the marginalized posterior PDFs that we obtain for these quantities, for the two separate populations as well as their sum.  Although the model is fit to the discrete multi-dimensional observations and not directly to the 1D histograms shown in Figure \ref{fig:cra2_scatter}, the fit shows generally good agreement with the histograms.  Moreover, our estimate of Crater 2's halflight radius remains in excellent agreement with the value T16 estimate from photometry alone.

\begin{table*}
\scriptsize
\centering
\caption{\scriptsize Summary of probability distribution functions for chemodynamical parameters---constant velocity dispersion model}
\begin{tabular}{@{}lllllllllll@{}}
\hline
parameter & prior & posterior & description\\
\hline
\smallskip
$\langle v_{\rm los}\rangle$ [km s$^{-1}$] & uniform between -500 and +500 & \gradientvmeanexpanded & mean velocity at center (Cra2)\\
\smallskip
$\log_{10}[\sigma^2_{v_{\rm los}} /(\mathrm{km^2 s^{-2}})]$ & uniform between -5 and +5 & \gradientvvarexpanded & velocity dispersion (Cra2)\\
\smallskip
$k_{v_{\rm los}}$ [km s$^{-1}$ arcmin$^{-1}$] & uniform between 0 and +10 & \gradientvgradexpanded & magnitude of maximum velocity gradient (Cra2)\\
\smallskip
$\theta_{v_{\rm los}}$ [deg.] & uniform between -180 and +180 & \gradientvthetaexpanded & direction of maximum velocity gradient (Cra2)\\
\smallskip
$\langle \feh\rangle $ [dex] & uniform between -5 and +1 & \gradientfehmeanexpanded & mean metallicity at center (Cra2)\\
\smallskip
$\log_{10}[\sigma^2_{\feh}]$ & uniform between -5 and +2 & \gradientzvarexpanded & metallicity dispersion (Cra2)\\
\smallskip
$k_Z$ [dex arcmin$^{-1}$] & uniform between -1 and +1 & \gradientfehgradexpanded & magnitude of metallicity gradient (Cra2)\\
\smallskip
$\log_{10}[\Sigma_{0,1}/(\mathrm{arcmin}^{-2})]$ & uniform between -10 and +10 & \gradientbigsigmaexpanded & 2D stellar density scale (Cra2)\\
\smallskip
$\log_{10}[R_{\rm h}/(\mathrm{arcmin})]$ & uniform between -1 and +3.5 & \gradientlogrsexpanded & 2D halflight radius (Cra2)\\
\smallskip
$\log_{10}[\Sigma_{0,2}/(\mathrm{arcmin}^{-2})]$ & uniform between -10 and +10 & \gradientbigsigmabexpanded & 2D stellar density (foreground)\\
\smallskip
$k_2 [\mathrm{arcmin}^{-1}]$ & uniform between 0 and +0.1 & \gradientbigsigmagradexpanded & gradient in 2D stellar density (foreground)\\
\smallskip
$\theta_2$ [deg.] & uniform between -180 and +180 & \gradientgradthetaexpanded & direction of gradient in 2D stellar density (foreground)\\
\smallskip
\smallskip
$\sigma_{v_{\rm los}}$ [km s$^{-1}$] && \gradientvdispexpanded & velocity dispersion (Cra2)\\
\smallskip
$\sigma_{\feh}$ [dex] && \gradientfehdispexpanded & metallicity dispersion (Cra2)\\
$R_{\rm h}$ [arcmin] && \gradientrsexpanded & 2D halflight radius (Cra2)\\
\hline
\end{tabular}
\\
\label{tab:cra2_gradient}
\end{table*}

\subsection{Velocity and Metallicity distributions}

\begin{figure}
  \includegraphics[width=3.5in,trim=0.25in 2.5in 1.0in 1.1in,clip]{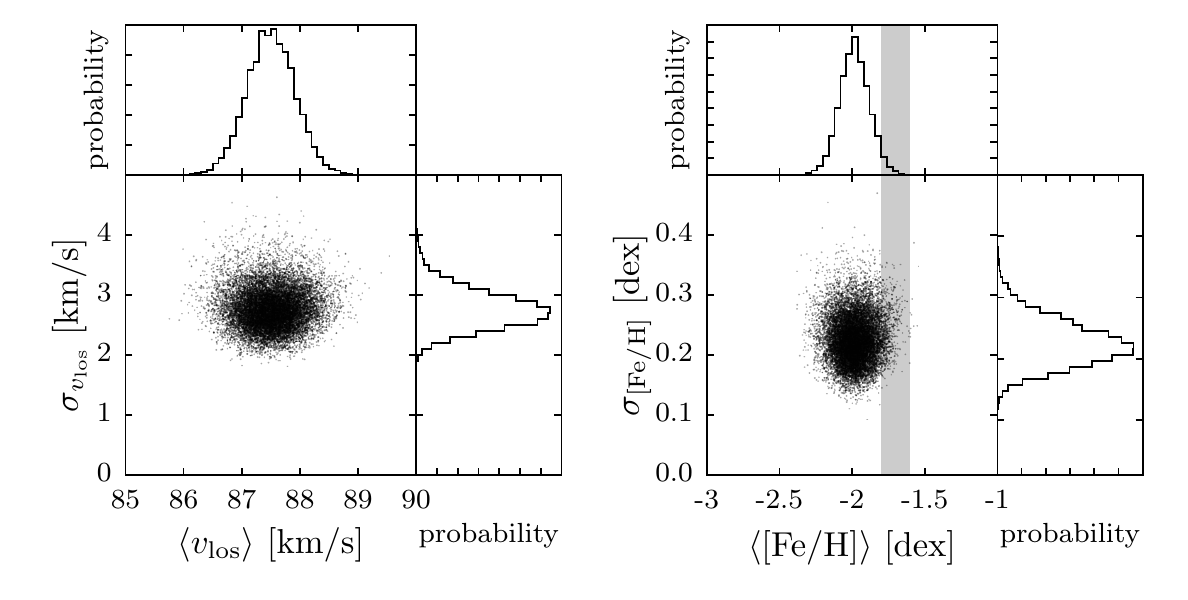}
  \caption{Samples drawn randomly from posterior PDFs for means and dispersions of velocity (left; solar rest frame) and metallicity (right) distributions for Crater 2.  Histograms display marginalized, 1D PDFs for each parameter.  In the right panel, the gray band represents the metallicity of $-1.7\pm 0.1$ that is estimated by fitting isochrones to the photometric data (T16; see Section \ref{subsec:metallicity}).}
  \label{fig:cra2_meansdispersions}
\end{figure}

Figure \ref{fig:cra2_meansdispersions} displays posterior PDFs we obtain for the means and dispersions of Crater 2's velocity and metallicity distributions.  For the member population we estimate a velocity dispersion of $\sigma_{v_{\los}}$$=$\gradientvdisp\ km s$^{-1}$ about a mean velocity of $\langle v_{\los}\rangle$=\gradientvmean\ km s$^{-1}$ in the solar rest frame.  While Crater 2's velocity distribution is extremely cold, it is well-resolved by our Hectochelle sample.  We find no evidence for a velocity gradient, placing only a (95\%) upper limit of $k_V\leq$ \gradientvgradupperlim\ km s$^{-1}$arcmin$^{-1}$.  We also resolve a metallicity dispersion of $\sigma_{\mathrm{[Fe/H]}}$$=$\gradientfehdisp\ dex about a mean of $\langle$[Fe/H]$\rangle=$ \gradientfehmean\ dex, with no evidence for a metallicity gradient.  

We also fit an alternative model in which any velocity gradient is attributed to the varying projection of Crater 2's systemic proper motion \citep{feast61,kaplinghat08,walker08}.  This model is identical to the one described above, except that the parameters $k_V$ and $\theta_V$ are replaced by proper motion coordinates, which specify the mean velocity at position $\vec{R}$ (see Appendix of \citet{walker08} for details).  We obtain estimates $\mu_{\alpha}=$\pmmualpha\ mas century$^{-1}$ and $\mu_{\delta}=$\pmmudelta\ mas century$^{-1}$ in the solar rest frame.  Thus our non-detection of a velocity gradient in the original model translates into a non-detection of departures from zero proper motion.

\subsection{Membership}
\label{subsec:membership}
Our mixture model lets us infer Crater 2's chemodynamical quantities without explicitly identifying member stars.  Nevertheless, it may be useful---e.g., in selecting targets for further observation---to identify likely members.  We use the posterior PDFs to compute, for each star with spectroscopic data, a posterior probability of membership\footnote{To save space, these probabilities are not listed in Table \ref{tab:cra2_table1}; however, they are included in the material that is made available in the online database.}:
\begin{equation}
  P(\mathrm{member}|\vec{R},V,Z,G,\Theta)=\frac{M}{M+N},
\end{equation}
where
\begin{eqnarray}
  M\equiv\Sigma_1(R|\Theta)P_1(V|\vec{R},\Theta)P_1(Z|R,\Theta)\hat{P}_1(G);\nonumber\\
  N\equiv \Sigma_2(R,\theta|\Theta)\hat{P}_2(V)\hat{P}_2(Z)\hat{P}_2(G).\hspace{0.55in}\nonumber
\end{eqnarray}
An estimate of the number of Crater 2 members within our spectroscopic sample is given by $\Sigma_{i=1}^{N_{\rm spec}}P(\mathrm{member}_i|\vec{R}_i,V_i,Z_i,G_i,\Theta)$$=$\nmem.  Red markers in Figures \ref{fig:cra2_cmd} and \ref{fig:cra2_scatter} identify the \nmemfifty\ stars for which $P(\mathrm{member}|\vec{R},V,Z,G,\Theta)>0.5$.  

Examining the top panel of Figure \ref{fig:cra2_scatter}, one probable member star (Cra2-224, with $P_{\mathrm{member}}|\vec{R},V,Z,G,\Theta)\sim 0.96$) has conspicuously high surface gravity ($\logg=5.29\pm 0.42$), which is more typical of the foreground contamination.  The reason for the large membership probability in this case is that the measured surface gravity has a relatively large error compared to the median error, and the measured velocity ($v_{\los}=82.9\pm 1.60$) and metallicity ([Fe/H]$=-2.02\pm 0.38$) are both close to the means we estimate for Crater 2.  In any case, our results for Crater 2 do not change if we discard this star altogether from our analysis.

\subsection{Comparison with Photometric Metallicity}
\label{subsec:metallicity}
Fitting isochrones to the ATLAS photometric data, T16 estimate that Crater 2's stellar population has age $10\pm 1$ Gyr and metallicity [Fe/H]$=-1.7\pm 0.1$.  Our spectroscopic metallicities indicate a lower mean metallicity of $\langle$[Fe/H]$\rangle$$=$\gradientfehmean.  Given this mild discrepancy, either the photometric estimate is systematically metal-rich or the spectroscopic metallicities that we obtain for individual stars are systematically metal-poor.  For two reasons, we consider the latter scenario to be more likely.  First, the colors and magnitudes of probable members in our spectroscopic data set show good agreement with the isochrone calculated for [Fe/H]=$-1.7$ (left panel of Figure \ref{fig:cra2_cmd}).  Second, our fits to the solar twilight spectra acquired during our Crater 2 observations (Section \ref{sec:obs}) return mean metallicity of $-0.58$ dex (Section \ref{sec:obs}).  While we treat this value as a zero-point offset and subtract it from all raw metallicity estimates, there is no guarantee that the same level of systematic error applies at all metallicities.  In any case, the disagreement between photometric and spectroscopic metallicities is significant only at the $\sim 2\sigma$ level.

Despite this likely systematic error, our spectroscopic metallicities remain useful for ranking stars by metallicity and for identifying metal-poor Crater 2 members amongst the Galactic foreground (Section \ref{subsec:membership}).  Furthermore, our spectroscopic detection of a significant  metallicity spread ($\sigma_{\mathrm{[Fe/H]}}$$=$\gradientfehdisp\ dex) is supported by visual inspection of the spectra for probable members.  

For example, the top two panels of Figure \ref{fig:cra2_spectra} display spectra for two probable members that have similar atmospheric parameters, but for which the best fitting templates yield [Fe/H] differences of $\sim$0.6 dex.  The more metal-rich star (top panel) displays stronger absorption features in general, including pronounced differences near the \ion{Ti}{2}/\ion{Fe}{1} features at 5227 \AA, a \ion{Ca}{1}/\ion{Fe}{1} feature at 5270 \AA, the \ion{Fe}{1} doublet at 5273 \AA, and an \ion{Fe}{2}/\ion{Cr}{1} feature at 5276 \AA.
%\begin{figure}[ht]
%\includegraphics[width=3.5in, trim=0in 2.5in 0.25in 2.5in,clip]{speccomp.pdf} 
%\caption{This figure compares the Hectochelle spectra of the Crater 2 stars S236 (black lines) and S119 (red lines).  The template fitting analysis (above)  found that these two stars likely have similar effective temperatures but differ in [Fe/H] by ~0.6 dex.  Significant changes in the line strengths of the Mg I b and Fe I features support the presence of an intrinsic metallicity spread in Crater 2.
%\label{fig:cra2_spectra}}
%\end{figure}

\subsection{Scaling Relations}
\label{subsec:scaling}

\begin{figure}[ht]
\includegraphics[width=3.5in, trim=0in 1.1in 3.25in 0in,clip]{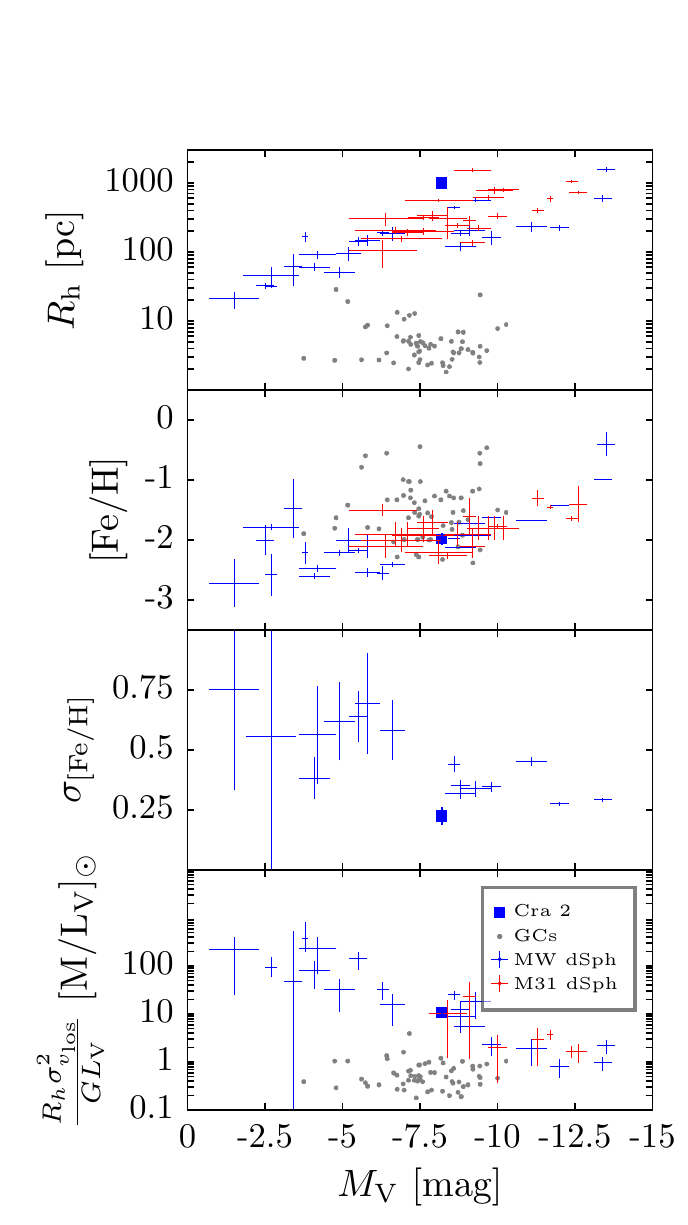} 
\caption{\textit{Top to bottom:} Size, mean metallicity, metallicity dispersion and dynamical mass-to-light ratio vs
  absolute magnitude, for Galactic globular clusters (black points) as
  well as dwarf spheroidal satellites of the Milky Way (blue points
  with errorbars) and M31 (red points with errorbars).  Quantities
  plotted for Crater 2 are adopted from \citet{torrealba16} and this work.  With the exception of metallicity dispersions, which are adopted from \citet{kirby08,kirby10,simon11,willman12} data for globular clusters and dSphs are adopted, respectively, from the catalog of \citet[2010 edition; we include only clusters with velocity dispersion measurements]{harris96} and the review of \citet{mcconnachie12}.
\label{fig:cra2_scaling1}}
\end{figure}

Figure \ref{fig:cra2_scaling1} places Crater 2 in the context of well-known scaling relations involving the Local Group's dwarf galaxies.  Crater 2's mean metallicity (regardless of whether we use the photometric estimate of [Fe/H]$=-1.7\pm 0.1$ or our spectroscopic estimate of $\langle \mathrm{[Fe/H]}\rangle $$=$\gradientfehmean) and (dimensional) dynamical mass-to-light ratio ($R_{\rm h}\sigma^2_V/L_VG$$=$\cramlratioscale\ $M_{\odot}/L_{\odot}$) are typical for dwarf galaxies of Crater 2's luminosity ($L_V=[1.5\pm 0.1]\times 10^5L_{V,\odot}$).  Nevertheless, Crater 2's metallicity dispersion is slightly smaller than those of other dwarfs of similar luminosity.  And, as already pointed out by T16, Crater 2 is a low-density outlier in the size-luminosity plane, with halflight radius ($R_h\sim 1.1$ kpc) nearly an order of magnitude larger than is typical for its luminosity (top panel of Figure \ref{fig:cra2_scaling1}).  

Our spectroscopic measurements reveal that Crater 2 is unusually large in another respect, regardless of its luminosity.  Figure \ref{fig:cra2_scaling2} shows scaling relations involving only the dynamical quantities of size and velocity dispersion.  While other dwarf galaxies with $R_{\rm h}\sim 1$ kpc have velocity dispersions of $\sim 10$ km s$^{-1}$ (top panel of Figure \ref{fig:cra2_scaling2}), Crater 2 is much colder, with $\sigma_{v_{\los}}$$=$ \gradientvdisp\ km s$^{-1}$.  Conversely, while there are other dwarf galaxies for which measured velocity dispersions are similarly cold, these dwarfs tend to be the the smallest `ultra-faints', with $R_{\rm h}\la 100$ pc (e.g., Segue 1, Segue 2, Leo IV, Leo V, Reticulum 2).  

\begin{figure}[ht]
\includegraphics[width=3.5in, trim=2.8in 0.5in 0in 2in,clip]{cra2_scaling.pdf} 
\caption{Velocity dispersion vs halflight radius, for the population
  of dwarf spheroidals plotted in Figure \ref{fig:cra2_scaling1} (excluding those for which the vertical errorbar is $> 100\%$).  Overplotted is the scaling relation $\mathcal{V}/(\mathrm{km\,s^{-1}})=\sqrt{R_{\rm h}/\mathrm{pc}}$, where $\mathcal{V}\equiv \sqrt{3}\sigma_{v_{\los}}$ \citep{mcgaugh07,walker10,walker14}.  
\label{fig:cra2_scaling2}}
\end{figure}

Moreover, among Local Group dwarf galaxies with resolved velocity dispersion measurements, Crater 2 is plausibly the coldest irrespective of other observables.  While \citet{koposov11} find evidence for a slightly colder (albeit with larger errorbars) sub-population in the Bootes I dwarf, with $\sigma_{v_{\los}}=2.4_{-0.5}^{+0.9}$ km s$^{-1}$, they measure an overall velocity dispersion of $\sigma_{v_{\los}}=4.6_{-0.6}^{+0.8}$ km s$^{-1}$ for Boo I.\footnote{We find no evidence for sub-populations within Crater 2.}  \citet{kirby13} and \citet{simon16} place (95\%) upper limits of $\sigma_{v_{\los}}<2.6$ km s$^{-1}$ and $\sigma_{v_{\los}}<1.5$ km s$^{-1}$ for Segue 2 and Tucana III, respectively, but given their median velocity errors of $3.7$ km s$^{-1}$ and $2.1$ km s$^{-1}$, they do not resolve these dispersions.  Also for Segue 2, \citet{belokurov09} report a resolved dispersion of $\sigma_{v_{\los}}=3.6_{-1.0}^{+1.7}$ km s$^{-1}$, based on a Hectochelle sample with median velocity error $0.6$ km s$^{-1}$.  Several dwarf satellites of M31 have velocity dispersions constrained to be $<3$ km s$^{-1}$, but all are unresolved \citep{collins13}.  Indeed, among known dwarfs, only Leo V has a (marginally) resolved velocity dispersion that is colder than Crater 2's.  Based on Hectochelle spectroscopy of Leo V, \citet{walker09c} report a marginally-resolved dispersion of $\sigma_{v_{\los}}=2.4_{-1.4}^{+2.4}$ km s$^{-1}$, most recently confirmed by \citet{collins16}, who report a marginally-resolved $\sigma_{v_{\los}}=2.3_{-1.6}^{+3.2}$ km s$^{-1}$.  While both estimates suggest Leo V is the colder galaxy, the available samples for Leo V are considerably smaller and the errorbars are considerably larger than we have for Crater 2.  

In any case, Crater 2 is the most securely cold outlier with respect to scaling relations involving size and velocity.  The simplest such relation is given by $\sigma^2_V\propto R_{\rm h}$, (\citealt{mcgaugh07}, \citealt{walker10}; top panel of Figure \ref{fig:cra2_scaling2}).  An equivalent relation is $\sigma^2_V/(R^2_{\rm h}G)\propto R_{\rm h}^{-1}$, which (dimensionally) takes the form of a `universal' mass-density profile (\citealt{walker09d}; bottom panel of Figure \ref{fig:cra2_scaling2}).  Other equivalent relations are $\sigma^2_V/(R_{\rm h}G)\propto 1$ \citep{kormendy85,donato09,salucci12}, and $\sigma^2_V/R_{\rm h}=$constant \citep{walker14}, implying constant scales for surface mass density and acceleration, respectively.  Previous deviations from these scaling relations, which have been shown to extend over all galactic scales, have plausibly been attributed to tidal stripping and/or to an offset between separate relations followed by the satellite populations of the Galaxy and M31 \citep{collins14}.  However, these arguments do not readily explain Crater 2's status as the most statistically significant outlier.  Not only is Crater 2 a satellite of the Galaxy and not M31, but its round morphology, lack of a velocity gradient, obeyance of the luminosity/metallicity relation, and large Galactocentric distance make it unlikely to have undergone significant tidal disruption.  In summary, Crater 2 is a diffuse, cold outlier in chemodynamical as well as structural scaling relations.  

\subsection{Dark Matter Halo}
\label{subsec:dm}

The available estimates of Crater 2's structural parameters and LOS velocity dispersion provide a crude approximation of the dynamical mass and mass-to-light ratio enclosed within a sphere that is centered on Crater 2 and has radius $r\sim R_{\rm h}$.  With the implicit assumptions of spherical symmetry,  dynamic equilibrium and negligible contribution of stellar binary motions to the observed velocity dispersion, the formula of \citet{walker09d} gives $M(R_{\rm h})\approx 5R_{\rm h}\sigma^2_V/(2G)$$=$\cramrhalf\ $M_{\odot}$, corresponding to a dynamical mass-to-light ratio of $\Upsilon(R_{\rm h})\approx 2M(R_{\rm h})/L_V$$=$\cramlratio\ $M_{\odot}/L_{V,\odot}$ and an equivalent circular velocity of $V_c(R_{\rm h})\equiv \sqrt{GM(R_{\rm h})/R_{\rm h}}$$=$\cravcirc\ km s$^{-1}$ at the halflight radius.  Under the same assumptions, the similar formula of \citet{wolf10} gives similar results, with $M(\frac{4}{3}R_{\rm h})\approx 4R_{\rm h}\sigma^2_V/G$$=$\cramwolf\ $M_{\odot}$, $\Upsilon(\frac{4}{3}R_{\rm h})\approx 2M(\frac{4}{3}R_{\rm h})/L_V$$=$\cramlwolf\ $M_{\odot}/L_{V,\odot}$ and $V_c(\frac{4}{3}R_{\rm h})\equiv \sqrt{GM(\frac{4}{3}R_{\rm h})/(\frac{4}{3}R_{\rm h})}$$=$\cravcircwolf km s$^{-1}$.  Thus despite its extreme coldness, Crater 2 is sufficiently large that---insofar as the stated assumptions are valid---its support against gravity requires a dominant dark matter halo.  

\subsubsection{Jeans Model}
\label{subsubsec:jeans}
The chemodynamical model described in previous sections assumes that the LOS velocity dispersion of Crater 2 members is independent of position.  In order to examine the dark matter content of Crater 2 in more detail, we shed this assumption by adopting a model in which the LOS velocity dispersion depends on position via the spherically-symmetric Jeans equation:
\begin{equation}
  \sigma^2_V(R|\Theta)=\frac{2}{\Sigma(R|\Theta)}\displaystyle \int_{R}^{\infty}\biggl (1-\beta(r)\frac{R^2}{r^2}\biggr ) \frac{\nu(r) \overline{v_r^2}(r)r}{\sqrt{r^2-R^2}}\deriv r,
  \label{eq:jeans}
\end{equation}
where \citep{binney82}
\begin{equation}
  \nu(r)\overline{v^2_r}(r)=\frac{1}{b(r)}\displaystyle\int_r^{\infty}b(s)\nu(s)\frac{GM(s)}{s^2}\deriv s.
  \label{eq:jeanssolution}
\end{equation}
Here, the function $b(r)\equiv b(r_1)\exp\bigl [2\int_{r_1}^{r}\beta_v(t)t^{-1}\deriv t\bigr ]$ is determined by the velocity dispersion anisotropy parameter $\beta_v(r)\equiv 1-\overline{v_{\theta}^2}(r)/\overline{v_r^2}(r)$, $\nu(r)$ is the 3D stellar number density profile (i.e., the deprojection of $\Sigma_1(R|\Theta)$), and $M(r)=M_{\rm DM}(r)+L_V(r)\Upsilon_{V,*}$ is the mass enclosed within radius $r$, which includes dark matter and stellar components (the enclosed stellar mass is the product of the enclosed luminosity, $L_V(r)$, and the stellar mass-to-light ratio, $\Upsilon_{V,*}$).  

Our adoption of a Plummer profile for $\Sigma(R|\Theta)$ implies the assumption $\nu(r)=\frac{3}{4}\Sigma_{0,1}R_{\rm h}^{-1}[1+R^2/R^2_{\rm h}]^{-5/2}$.  We assume the dark matter halo has density profile of the form \citep{zhao96}
\begin{equation}
  \rho_{\rm DM}(r)=\rho_s\biggl (\frac{r}{r_s}\biggr )^{-\gamma}\biggl [1+\biggl (\frac{r}{r_s}\biggr )^{\alpha}\biggr ]^{(\gamma-\beta)/\alpha},
\end{equation}
which specifies the enclosed dark mass profile $M_{\rm DM}(r)=4\pi\int_{0}^{r}s^2\rho_{\rm DM}(s)\deriv s$.  

With these assumptions, we replace the free parameter $\sigma_{v_{\los}}$ with the five free parameters that specify $\rho_{\rm DM}(r)$, a free parameter for the stellar mass-to-light ratio, and an additional parameter that specifies $\beta_v$, which we assume to be constant.  We adopt the same broad priors on halo parameters and anisotropy as described by \citet{geringer-sameth15b}.  For the stellar mass-to-light ratio, we adopt a prior that is uniform between $-1\leq \log_{10}[\Upsilon_{*,V}/(M_{\odot}L_{\odot}^{-1})]$.  Unlike in previous work, our Jeans model is now embedded within the mixture model, specifying the velocity dispersion of Crater 2 members as a function of position.  

\begin{figure}[ht]
\includegraphics[width=3.5in, trim=0in 2.65in 3in 0.5in,clip]{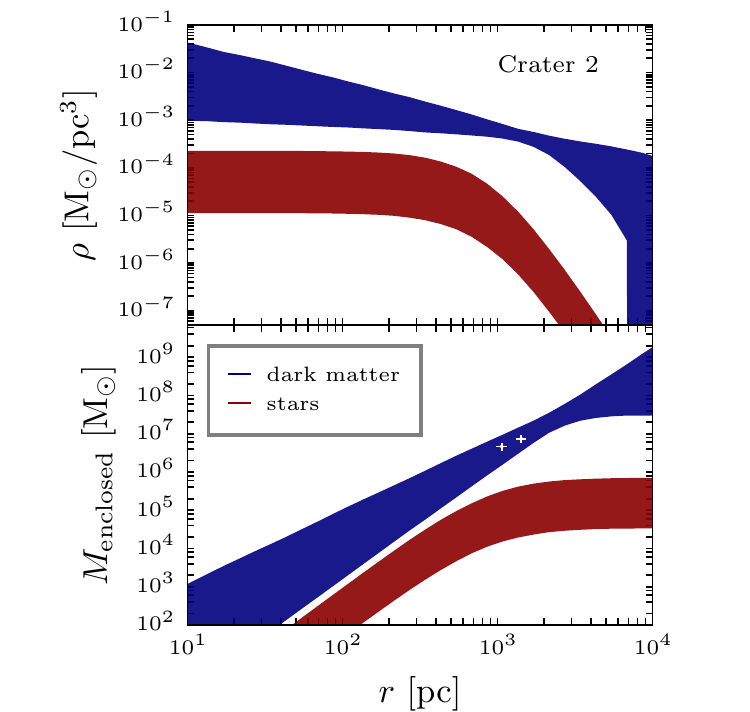} 
\caption{Density and enclosed-mass profiles for Crater 2's stellar and dark matter components, from our mixture model based on the Jeans equation (Section \ref{subsec:dm}).  Colored bands represent 68\% credibility intervals at each radius.  Crosses represent estimates of masses enclosed within spheres of radius $R_{\rm h}$ \citep{walker09d} and $\frac{4}{3}R_{\rm h}$ \citep{wolf10}.
\label{fig:cra2jeans_profiles1}}
\end{figure}

Figure \ref{fig:cra2jeans_profiles1} displays 68\% credibility intervals that we obtain for stellar and dark matter profiles for enclosed-mass and density.  Again despite Crater 2's extremely cold velocity dispersion, and again subject to the validity of the assumptions of dynamic equilibrium and negligible contamination from binary stars, we find that dark matter dominates Crater 2's gravitational potential at all radii, with the dark matter density exceeding that contributed by stars by two orders of magnitude even at $r=0$.  Moreover, at radii $r\sim R_{\rm h}$, the results of our Jeans model stand in excellent agreement with the mass estimators of \citet{walker09d} and \citet[bottom panel of Figure \ref{fig:cra2jeans_profiles1}]{wolf10}.  

\subsubsection{Astrophysical `$J$'-factor for Dark Matter Searches}
\label{subsubsec:J}
Although this analysis suggests that Crater 2's gravitational potential is dominated by dark matter even at its center, the overall amount of dark matter in Crater 2---as constrained, say, within the halflight radius of $r=R_{\rm h}\sim 1$ kpc---is modest relative to what has been estimated for other Galactic dwarf spheroidals at similar radii \citep{strigari08,walker09d,wolf10}.  Combined with a relatively large distance of $\sim 120$ kpc, these attributes make Crater 2 a poor target in the search for indirect evidence of dark matter particle interactions---e.g., annihilation to gamma-rays.  Using the dark matter density profile estimated above, we use the method of \citet{geringer-sameth15b} to compute the profile $\deriv J/\deriv\Omega\equiv \int_0^{\infty}\rho^2(l)\deriv l$, where $l$ increases along the line of sight and the `J-factor' $J(\theta)\equiv \int_0^{\theta}2\pi\sin(\theta')\deriv\theta'\deriv J(\theta')/\deriv\Omega$ is proportional to the flux of annihilation photons for a given particle physics model.  At $\theta=1.4^{\circ}$, the angle corresponding to the projected radius of the outermost spectroscopic member, we obtain $\log_{10}[J/(\mathrm{GeV}^2\mathrm{cm}^{-5})]=15.7\pm 0.25$.  For comparison, the most attractive dwarf-galactic targets for dark matter searches have  $\log_{10}[J/(\mathrm{GeV}^2\mathrm{cm}^{-5})]\ga 19$, even when evaluated at smaller angles \citep{geringer-sameth15b,bonnivard15}.  

\begin{figure}[ht]
\includegraphics[width=3.5in, trim=3in 0.2in 0.2in 3.75in,clip]{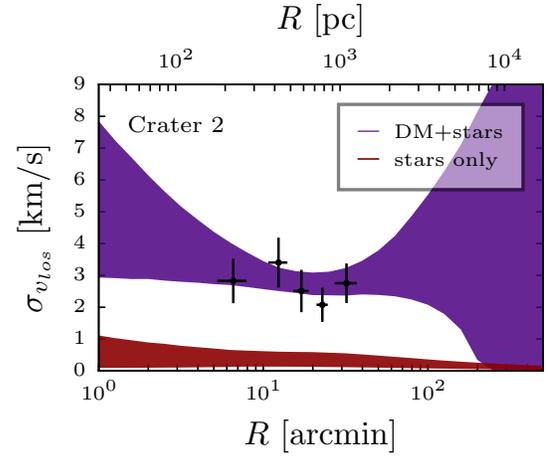} 
\caption{Projected velocity dispersion profile of Crater 2.  The purple colored band indicates the 68\% credibility interval at each projected radius, from our mixture model based on the Jeans equation (Section \ref{subsec:dm}).  Black crosses indicate the binned velocity dispersion profile estimated using the resulting membership probabilities as weights (note: the model represented by the purple shaded region was fit to unbinned photometric and spectroscopic data).  The red colored band indicates the velocity dispersion profile expected in the case that Crater 2 lacks a dark matter halo and is bound only by the self-gravity of its stars, allowing for stellar mass-to-light ratios in the range $-1\leq \log_{10}[\Upsilon_*/(M_{\odot}/L_{\odot,V})]\leq 1$.  
\label{fig:cra2jeans_profiles2}}
\end{figure}

Finally, Figure \ref{fig:cra2jeans_profiles2} shows the LOS velocity dispersion profile of Crater 2, which we calculate by weighting all spectroscopic data points by the posterior probability of membership, given the data and our mixture model based on the Jeans equation.  Where it is constrained by data, the velocity dispersion profile is approximately flat, as is the case for nearly all well-studied dwarf galaxies \citep{walker07b}.  The purple band indicates 68\% credibility intervals on the velocity dispersion as a function of radius, calculated directly from the Jeans model described above.  The red band indicates the velocity dispersion profile calculated from the Jeans equation, but in the case that there is no dark matter halo---ie., the gravitational potential arises only from the self-gravity of Crater 2 stars, for which we continue to allow stellar mass-to-light ratios in the broad range $-1\leq \log_{10}[\Upsilon_*/(M_{\odot}/L_{\odot})]\leq 1$.  Compared to the original Jeans model that allows for a dark matter halo, this version without dark matter is disfavored overwhelmingly, as quantified by a Bayesian evidence ratio of $e^{106}$.

\section{Summary \& Discussion}

Table \ref{tab:summarytable} summarizes the observed properties of Crater 2, listing the structural parameters measured by T16 along with chemodynamical parameters measured from our Hectochelle spectroscopy.  While Crater 2's metallicity and dynamical mass-to-light ratio are similar to those of other dwarf galaxies of similar luminosity, its large size and cold velocity dispersion make Crater 2 an extremely low outlier in terms of both stellar surface density and dynamical mass-density.  
\begin{table*}
  \begin{centering}
  \scriptsize
  \caption{Summary of observed photometric and spectroscopic properties for Crater 2}
  \begin{tabular}{@{}lllllllllll@{}}
    \hline
    quantity& value &description&reference\\
    \hline
    $\alpha_{\rm J2000}$&11:39:31&R.A. at center&T16$^{1}$&\\
    $\delta_{\rm J2000}$&$-$18:24:47&Dec. at center&T16\\
    $l$ [deg]&\cragalacticl&Galactic longitude&T16\\
    $b$ [deg]&\cragalacticb&Galactic latitude&T16\\
    $m-M$ [mag]&$20.35\pm 0.02$&distance modulus&T16\\
    $D$ [kpc]&$117.5\pm 1.1$&distance from Sun&T16\\
    $\tau$ [Gyr]&$10\pm 1$&age&T16\\
    $\mathrm{[Fe/H]}$&$-1.7\pm 0.1$&photometric metallicity (from isochrone fitting)&T16\\
    $M_{\rm V}$ [mag]&$-8.2\pm 0.1$&absolute magnitude&T16\\
    $R_{\rm h}$ [arcmin]&$31.2\pm 2.5$&projected halflight radius$^{2}$&T16\\
    $R_{\rm h}$ [pc]&$1066\pm 84$&projected halflight radius$^{2}$&T16\\
\\
    $v_{\rm los}$ [km s$^{-1}$]& \gradientvmean & mean line-of-sight velocity, solar rest frame&this work\\
    $v_{\rm los}$ [km s$^{-1}$]&\cravgrf & mean line-of-sight velocity, Galactic rest frame$^{2}$&this work\\
    $\sigma_{v_{\los}}$ [km s$^{-1}$]&\gradientvdisp& internal velocity dispersion&this work\\
    $k_{v_{\los}}$ [km s$^{-1}$ arcmin$^{-1}$]&$<$ \gradientvgradupperlim & velocity gradient&this work\\
    $\theta_{v_{\los}}$ [deg]& \nodata & PA of velocity gradient&this work\\
    $\mu_{\alpha}$ [mas/century]&\pmmualpha & R.A. proper motion (solar rest frame)& this work\\
    $\mu_{\delta}$ [mas/century]&\pmmudelta & Dec. proper motion (solar rest frame)&this work\\
    $\langle \feh\rangle$ [dex]& \gradientfehmean &mean spectroscopic metallicity&this work\\
    $\sigma_{\feh}$ [dex]& \gradientfehdisp &spectroscopic metallicity dispersion&this work\\
    $k_Z$ [dex arcmin$^{-1}$]&\gradientfehgrad  &spectroscopic metallicity gradient&this work\\
    $M(R_{\rm h})$ [$\mathrm{M}_{\odot}$]& \cramrhalf & dynamical mass$^{3}$ enclosed within $r=R_{\rm h}$&this work\\
    $V_c(R_{\rm h})$ [km s$^{-1}$]& \cravcirc & circular velocity$^{4}$ at $r=R_{\rm h}$&this work\\
    $\Upsilon$ [$\mathrm{M}_{\odot}/L_{V,\odot}$]& \cramlratio &dynamical mass-to-light ratio$^5$ within $R_{\rm h}$&this work\\
    $\log_{10}[J/(\mathrm{GeV}^2\mathrm{cm}^{-5})]$&$15.7\pm 0.25$&$J$-factor for dark matter annihilation&this work\\
    \hline
    \multicolumn{5}{l}{$^{1}$ \citet{torrealba16}}\\
    \multicolumn{5}{l}{$^{2}$ calculated using the solar motion measured by \citet{schonrich10}}\\
    \multicolumn{5}{l}{$^{3}$ $M(R_{\rm h})\approx 5R_{\rm h}\sigma_{v_{\los}}^2/(2G)$; assumes equilibrium, negligible binary stars}\\
    \multicolumn{5}{l}{$^{4}$ $V_c(R_{\rm h})\equiv\sqrt{GM(R_{\rm h})/R_{\rm h}}$}\\
    \multicolumn{5}{l}{$^{5}$ $\Upsilon\approx 2M(R_{\rm h})/L_V$}\\
  \end{tabular}
  \label{tab:summarytable}
\end{centering}
\end{table*}

For an object of Crater 2's luminosity and size, the self-gravity of the stars alone generates a line-of-sight velocity dispersion of just $< 1$ km s$^{-1}$.  Taken at face value, even velocity dispersions as cold as $\sim 2-3$ km s$^{-1}$ imply a dominant dark matter component, as we have found.  However, given the ability of unresolved binary-star orbital motions alone to generate velocity dispersions of the observed magnitude (\citealt{edo96,hargreaves96b,mcconnachie10}, Spencer et al., in preparation), the current dynamical evidence for dark matter in Crater 2 is somewhat precarious.  Robust estimates of Crater 2's dark matter content will require repeat spectroscopic observations to detect and quantify the properties of its binary stars.  To the extent that binary stars contribute to the velocity dispersion that we have measured, Crater 2 is intrinsically even colder and less dense than indicated above.

It will be up to a combination of numerical simulations and various dark matter models to interpret these results in the context of a galaxy formation theory.  For example, star formation and subsequent supernova winds can give up gravitational energy to standard `cold' dark matter (CDM), lowering halo concentration as dark matter expands non-adiabatically \citep[and references therein]{pontzen14}.  For example, examining the simulated dark matter halos of Milky-Way-dwarf-satellite analogs down to Crater 2's luminosity, \citet{wetzel16} show how this sort of process drops circular velocities by as much as $\sim 25$ km s$^{-1}$ to $\sim 10$ km s$^{-1}$ at radii of 1 kpc.  However, the value that we estimate for Crater 2 would be off the lower end of the relation between $V_c$ vs $r$ plotted in Figure 2 of \citet{wetzel16}.  It remains to be seen whether the standard CDM model can be expected to produce objects like Crater 2 around galaxies like the Milky Way.

Alternatively, it may turn out that Crater 2 is more naturally produced in models that ascribe more exotic properties to the dark matter itself.  For example, non-gravitational self-scattering of dark matter particles can flatten the central density `cusps' that characterize halos formed in CDM cosmological simulations \citep[e.g., ][]{spergel00,loeb11}, perhaps lowering central dark matter densities to the level we infer for Crater 2.  Another possibility is that the thermal free-streaming of sufficiently `warm' dark matter---e.g., sterile neutrinos \citep{dodelson94}---might prevent the formation of dense cusps in the first place \citep{bode01}.  The same effect might also result from quantum pressure associated with the de Broglie wavelength of particles---e.g., axions---giving rise to light scalar fields \citep{marsh15}.  

Finally, we note that \citet{mcgaugh16b} has recently calculated the velocity dispersion expected for Crater 2 under the hypothesis of Modified Newtonian Dynamics (MOND; \citealt{milgrom83}), which fits galactic rotation curves not by invoking dark matter, but rather by modifying gravity in the regime of low accelerations, $a\equiv GM_{\rm b}(r)/r^2\ll a_0$, where $M_{\rm b}(r)$ is the baryonic mass interior to radius $r$ and $a_0\sim 1.2\times 10^{-10}$ m s$^{-2}$.  Assuming spherical symmetry, isotropic velocity dispersions and adopting stellar mass-to-light ratio $\Upsilon_*=2_{-1}^{+2} M_{\odot}/L_{\odot}$, \citet{mcgaugh16b} uses the formula of \citet{mcgaugh13} to predict for Crater 2 a velocity dispersion of $\sigma_{v_{\los}}\approx 4$ km s$^{-1}$ in the `deep-MOND' limit that ignores the external acceleration field of the Milky Way.  Accounting for the external field lowers the prediction to $\sigma_{v_{\los}}=2.1_{-0.6}^{+0.9}$ km s$^{-1}$, where the errorbars propagate uncertainty in the adopted $\Upsilon_*$.  This prediction is consistent with the otherwise-unexpectedly cold velocity dispersion that we measure.

Of course, no model can be tested definitively based on what it predicts (or does not predict) for Crater 2 alone.  But by extending the range of properties exhibited by the Milky Way's satellites, Crater 2 gives that population more leverage to distinguish amongst various models.  Moreover, the discovery of Crater 2 at current surface brightness detection limits gives reason to be optimistic that the next generation of sky surveys will uncover even more extreme objects.

\acknowledgments
M.G.W. is supported by National Science Foundation grants AST-1313045 and AST-1412999.  M.M. is supported by NSF grant AST-1312997.  E.W.O. is supported by NSF grant AST-1313006.  C.I.J. gratefully acknowledges support from
the Clay Fellowship, administered by the Smithsonian Astrophysical Observatory.  S.K. thanks the United Kingdom Science and Technology Council (STFC) for the award of Ernest Rutherford fellowship (grant number ST/N004493/1).  The research leading to these results has received funding from the European Research Council under the European Union's Seventh Framework Programme (FP/2007-2013)/ERC Grant Agreement no. 308024.

\clearpage

%%TABLES

%\bibliography{ref}

\begin{thebibliography}{71}
\expandafter\ifx\csname natexlab\endcsname\relax\def\natexlab#1{#1}\fi

\bibitem[{{Battaglia et al.}(2006)}]{battaglia06}
{Battaglia et al.} 2006, \aap, 459, 423

\bibitem[{{Bechtol} {et~al.}(2015){Bechtol}, {Drlica-Wagner}, {Balbinot},
  {Pieres}, {Simon}, {Yanny}, {Santiago}, {Wechsler}, {Frieman}, {Walker},
  {Williams}, {Rozo}, {Rykoff}, {Queiroz}, {Luque}, {Benoit-L{\'e}vy},
  {Tucker}, {Sevilla}, {Gruendl}, {da Costa}, {Fausti Neto}, {Maia}, {Abbott},
  {Allam}, {Armstrong}, {Bauer}, {Bernstein}, {Bernstein}, {Bertin}, {Brooks},
  {Buckley-Geer}, {Burke}, {Carnero Rosell}, {Castander}, {Covarrubias},
  {D'Andrea}, {DePoy}, {Desai}, {Diehl}, {Eifler}, {Estrada}, {Evrard},
  {Fernandez}, {Finley}, {Flaugher}, {Gaztanaga}, {Gerdes}, {Girardi},
  {Gladders}, {Gruen}, {Gutierrez}, {Hao}, {Honscheid}, {Jain}, {James},
  {Kent}, {Kron}, {Kuehn}, {Kuropatkin}, {Lahav}, {Li}, {Lin}, {Makler},
  {March}, {Marshall}, {Martini}, {Merritt}, {Miller}, {Miquel}, {Mohr},
  {Neilsen}, {Nichol}, {Nord}, {Ogando}, {Peoples}, {Petravick}, {Plazas},
  {Romer}, {Roodman}, {Sako}, {Sanchez}, {Scarpine}, {Schubnell}, {Smith},
  {Soares-Santos}, {Sobreira}, {Suchyta}, {Swanson}, {Tarle}, {Thaler},
  {Thomas}, {Wester}, {Zuntz}, \& {The DES Collaboration}}]{des15}
{Bechtol}, K., {Drlica-Wagner}, A., {Balbinot}, E., {Pieres}, A., {Simon},
  J.~D., {Yanny}, B., {Santiago}, B., {Wechsler}, R.~H., {Frieman}, J.,
  {Walker}, A.~R., {Williams}, P., {Rozo}, E., {Rykoff}, E.~S., {Queiroz}, A.,
  {Luque}, E., {Benoit-L{\'e}vy}, A., {Tucker}, D., {Sevilla}, I., {Gruendl},
  R.~A., {da Costa}, L.~N., {Fausti Neto}, A., {Maia}, M.~A.~G., {Abbott}, T.,
  {Allam}, S., {Armstrong}, R., {Bauer}, A.~H., {Bernstein}, G.~M.,
  {Bernstein}, R.~A., {Bertin}, E., {Brooks}, D., {Buckley-Geer}, E., {Burke},
  D.~L., {Carnero Rosell}, A., {Castander}, F.~J., {Covarrubias}, R.,
  {D'Andrea}, C.~B., {DePoy}, D.~L., {Desai}, S., {Diehl}, H.~T., {Eifler},
  T.~F., {Estrada}, J., {Evrard}, A.~E., {Fernandez}, E., {Finley}, D.~A.,
  {Flaugher}, B., {Gaztanaga}, E., {Gerdes}, D., {Girardi}, L., {Gladders}, M.,
  {Gruen}, D., {Gutierrez}, G., {Hao}, J., {Honscheid}, K., {Jain}, B.,
  {James}, D., {Kent}, S., {Kron}, R., {Kuehn}, K., {Kuropatkin}, N., {Lahav},
  O., {Li}, T.~S., {Lin}, H., {Makler}, M., {March}, M., {Marshall}, J.,
  {Martini}, P., {Merritt}, K.~W., {Miller}, C., {Miquel}, R., {Mohr}, J.,
  {Neilsen}, E., {Nichol}, R., {Nord}, B., {Ogando}, R., {Peoples}, J.,
  {Petravick}, D., {Plazas}, A.~A., {Romer}, A.~K., {Roodman}, A., {Sako}, M.,
  {Sanchez}, E., {Scarpine}, V., {Schubnell}, M., {Smith}, R.~C.,
  {Soares-Santos}, M., {Sobreira}, F., {Suchyta}, E., {Swanson}, M.~E.~C.,
  {Tarle}, G., {Thaler}, J., {Thomas}, D., {Wester}, W., {Zuntz}, J., \& {The
  DES Collaboration}. 2015, \apj, 807, 50

\bibitem[{{Belokurov} {et~al.}(2009){Belokurov}, {Walker}, {Evans}, {Gilmore},
  {Irwin}, {Mateo}, {Mayer}, {Olszewski}, {Bechtold}, \&
  {Pickering}}]{belokurov09}
{Belokurov}, V., {Walker}, M.~G., {Evans}, N.~W., {Gilmore}, G., {Irwin},
  M.~J., {Mateo}, M., {Mayer}, L., {Olszewski}, E., {Bechtold}, J., \&
  {Pickering}, T. 2009, ArXiv:0903.0818

\bibitem[{{Belokurov et al.}(2007)}]{belokurov07}
{Belokurov et al.} 2007, \apj, 654, 897

\bibitem[{{Binney} \& {Mamon}(1982)}]{binney82}
{Binney}, J., \& {Mamon}, G.~A. 1982, \mnras, 200, 361

\bibitem[{{Bode} {et~al.}(2001){Bode}, {Ostriker}, \& {Turok}}]{bode01}
{Bode}, P., {Ostriker}, J.~P., \& {Turok}, N. 2001, \apj, 556, 93

\bibitem[{{Bonnivard} {et~al.}(2015){Bonnivard}, {Combet}, {Daniel}, {Funk},
  {Geringer-Sameth}, {Hinton}, {Maurin}, {Read}, {Sarkar}, {Walker}, \&
  {Wilkinson}}]{bonnivard15}
{Bonnivard}, V., {Combet}, C., {Daniel}, M., {Funk}, S., {Geringer-Sameth}, A.,
  {Hinton}, J.~A., {Maurin}, D., {Read}, J.~I., {Sarkar}, S., {Walker}, M.~G.,
  \& {Wilkinson}, M.~I. 2015, \mnras, 453, 849

\bibitem[{{Caldwell} {et~al.}(2009){Caldwell}, {Harding}, {Morrison}, {Rose},
  {Schiavon}, \& {Kriessler}}]{caldwell09}
{Caldwell}, N., {Harding}, P., {Morrison}, H., {Rose}, J.~A., {Schiavon}, R.,
  \& {Kriessler}, J. 2009, \aj, 137, 94

\bibitem[{{Collins} {et~al.}(2013){Collins}, {Chapman}, {Rich}, {Ibata},
  {Martin}, {Irwin}, {Bate}, {Lewis}, {Pe{\~n}arrubia}, {Arimoto}, {Casey},
  {Ferguson}, {Koch}, {McConnachie}, \& {Tanvir}}]{collins13}
{Collins}, M.~L.~M., {Chapman}, S.~C., {Rich}, R.~M., {Ibata}, R.~A., {Martin},
  N.~F., {Irwin}, M.~J., {Bate}, N.~F., {Lewis}, G.~F., {Pe{\~n}arrubia}, J.,
  {Arimoto}, N., {Casey}, C.~M., {Ferguson}, A.~M.~N., {Koch}, A.,
  {McConnachie}, A.~W., \& {Tanvir}, N. 2013, \apj, 768, 172

\bibitem[{{Collins} {et~al.}(2014){Collins}, {Chapman}, {Rich}, {Ibata},
  {Martin}, {Irwin}, {Bate}, {Lewis}, {Pe{\~n}arrubia}, {Arimoto}, {Casey},
  {Ferguson}, {Koch}, {McConnachie}, \& {Tanvir}}]{collins14}
---. 2014, \apj, 783, 7

\bibitem[{{Collins} {et~al.}(2016){Collins}, {Tollerud}, {Sand}, {Bonaca},
  {Willman}, \& {Strader}}]{collins16}
{Collins}, M.~L.~M., {Tollerud}, E.~J., {Sand}, D.~J., {Bonaca}, A., {Willman},
  B., \& {Strader}, J. 2016, ArXiv:1608.05710

\bibitem[{{Di Cintio} {et~al.}(2014){Di Cintio}, {Brook}, {Dutton},
  {Macci{\`o}}, {Stinson}, \& {Knebe}}]{dicintio14}
{Di Cintio}, A., {Brook}, C.~B., {Dutton}, A.~A., {Macci{\`o}}, A.~V.,
  {Stinson}, G.~S., \& {Knebe}, A. 2014, \mnras, 441, 2986

\bibitem[{{Dodelson} \& {Widrow}(1994)}]{dodelson94}
{Dodelson}, S., \& {Widrow}, L.~M. 1994, Physical Review Letters, 72, 17

\bibitem[{{Donato} {et~al.}(2009){Donato}, {Gentile}, {Salucci}, {Frigerio
  Martins}, {Wilkinson}, {Gilmore}, {Grebel}, {Koch}, \& {Wyse}}]{donato09}
{Donato}, F., {Gentile}, G., {Salucci}, P., {Frigerio Martins}, C.,
  {Wilkinson}, M.~I., {Gilmore}, G., {Grebel}, E.~K., {Koch}, A., \& {Wyse}, R.
  2009, \mnras, 397, 1169

\bibitem[{{Dotter} {et~al.}(2008){Dotter}, {Chaboyer}, {Jevremovi{\'c}},
  {Kostov}, {Baron}, \& {Ferguson}}]{dotter08}
{Dotter}, A., {Chaboyer}, B., {Jevremovi{\'c}}, D., {Kostov}, V., {Baron}, E.,
  \& {Ferguson}, J.~W. 2008, \apjs, 178, 89

\bibitem[{{Feast} {et~al.}(1961){Feast}, {Thackeray}, \& {Wesselink}}]{feast61}
{Feast}, M.~W., {Thackeray}, A.~D., \& {Wesselink}, A.~J. 1961, \mnras, 122,
  433

\bibitem[{{Feroz} \& {Hobson}(2008)}]{feroz08}
{Feroz}, F., \& {Hobson}, M.~P. 2008, \mnras, 384, 449

\bibitem[{{Feroz} {et~al.}(2009){Feroz}, {Hobson}, \& {Bridges}}]{feroz09}
{Feroz}, F., {Hobson}, M.~P., \& {Bridges}, M. 2009, \mnras, 398, 1601

\bibitem[{{Geringer-Sameth} {et~al.}(2015){Geringer-Sameth}, {Koushiappas}, \&
  {Walker}}]{geringer-sameth15b}
{Geringer-Sameth}, A., {Koushiappas}, S.~M., \& {Walker}, M. 2015, \apj, 801,
  74

\bibitem[{{Hargreaves} {et~al.}(1996){Hargreaves}, {Gilmore}, \&
  {Annan}}]{hargreaves96b}
{Hargreaves}, J.~C., {Gilmore}, G., \& {Annan}, J.~D. 1996, \mnras, 279, 108

\bibitem[{{Harris}(1996)}]{harris96}
{Harris}, W.~E. 1996, \aj, 112, 1487

\bibitem[{{Kaplinghat} \& {Strigari}(2008)}]{kaplinghat08}
{Kaplinghat}, M., \& {Strigari}, L.~E. 2008, \apjl, 682, L93

\bibitem[{{Kirby} {et~al.}(2013){Kirby}, {Cohen}, {Guhathakurta}, {Cheng},
  {Bullock}, \& {Gallazzi}}]{kirby13}
{Kirby}, E.~N., {Cohen}, J.~G., {Guhathakurta}, P., {Cheng}, L., {Bullock},
  J.~S., \& {Gallazzi}, A. 2013, \apj, 779, 102

\bibitem[{{Kirby} {et~al.}(2010){Kirby}, {Guhathakurta}, {Simon}, {Geha},
  {Rockosi}, {Sneden}, {Cohen}, {Sohn}, {Majewski}, \& {Siegel}}]{kirby10}
{Kirby}, E.~N., {Guhathakurta}, P., {Simon}, J.~D., {Geha}, M.~C., {Rockosi},
  C.~M., {Sneden}, C., {Cohen}, J.~G., {Sohn}, S.~T., {Majewski}, S.~R., \&
  {Siegel}, M. 2010, \apjs, 191, 352

\bibitem[{{Kirby} {et~al.}(2008){Kirby}, {Simon}, {Geha}, {Guhathakurta}, \&
  {Frebel}}]{kirby08}
{Kirby}, E.~N., {Simon}, J.~D., {Geha}, M., {Guhathakurta}, P., \& {Frebel}, A.
  2008, \apjl, 685, L43

\bibitem[{{Koposov} {et~al.}(2015){Koposov}, {Belokurov}, {Torrealba}, \&
  {Evans}}]{koposov15}
{Koposov}, S.~E., {Belokurov}, V., {Torrealba}, G., \& {Evans}, N.~W. 2015,
  \apj, 805, 130

\bibitem[{{Koposov} {et~al.}(2011){Koposov}, {Gilmore}, {Walker}, {Belokurov},
  {Wyn Evans}, {Fellhauer}, {Gieren}, {Geisler}, {Monaco}, {Norris}, {Okamoto},
  {Pe{\~n}arrubia}, {Wilkinson}, {Wyse}, \& {Zucker}}]{koposov11}
{Koposov}, S.~E., {Gilmore}, G., {Walker}, M.~G., {Belokurov}, V., {Wyn Evans},
  N., {Fellhauer}, M., {Gieren}, W., {Geisler}, D., {Monaco}, L., {Norris},
  J.~E., {Okamoto}, S., {Pe{\~n}arrubia}, J., {Wilkinson}, M., {Wyse},
  R.~F.~G., \& {Zucker}, D.~B. 2011, \apj, 736, 146

\bibitem[{{Koposov et al.}(2008)}]{koposov08}
{Koposov et al.} 2008, \apj, 686, 279

\bibitem[{{Kormendy}(1985)}]{kormendy85}
{Kormendy}, J. 1985, \apj, 295, 73

\bibitem[{{Kormendy} \& {Bender}(2012)}]{kormendy12}
{Kormendy}, J., \& {Bender}, R. 2012, \apjs, 198, 2

\bibitem[{{Lee} {et~al.}(2008{\natexlab{a}}){Lee}, {Beers}, {Sivarani},
  {Allende Prieto}, {Koesterke}, {Wilhelm}, {Re Fiorentin}, {Bailer-Jones},
  {Norris}, {Rockosi}, {Yanny}, {Newberg}, {Covey}, {Zhang}, \& {Luo}}]{lee08a}
{Lee}, Y.~S., {Beers}, T.~C., {Sivarani}, T., {Allende Prieto}, C.,
  {Koesterke}, L., {Wilhelm}, R., {Re Fiorentin}, P., {Bailer-Jones}, C.~A.~L.,
  {Norris}, J.~E., {Rockosi}, C.~M., {Yanny}, B., {Newberg}, H.~J., {Covey},
  K.~R., {Zhang}, H.-T., \& {Luo}, A.-L. 2008{\natexlab{a}}, \aj, 136, 2022

\bibitem[{{Lee} {et~al.}(2008{\natexlab{b}}){Lee}, {Beers}, {Sivarani},
  {Johnson}, {An}, {Wilhelm}, {Allende Prieto}, {Koesterke}, {Re Fiorentin},
  {Bailer-Jones}, {Norris}, {Yanny}, {Rockosi}, {Newberg}, {Cudworth}, \&
  {Pan}}]{lee08b}
{Lee}, Y.~S., {Beers}, T.~C., {Sivarani}, T., {Johnson}, J.~A., {An}, D.,
  {Wilhelm}, R., {Allende Prieto}, C., {Koesterke}, L., {Re Fiorentin}, P.,
  {Bailer-Jones}, C.~A.~L., {Norris}, J.~E., {Yanny}, B., {Rockosi}, C.,
  {Newberg}, H.~J., {Cudworth}, K.~M., \& {Pan}, K. 2008{\natexlab{b}}, \aj,
  136, 2050

\bibitem[{{Loeb} \& {Weiner}(2011)}]{loeb11}
{Loeb}, A., \& {Weiner}, N. 2011, Physical Review Letters, 106, 171302

\bibitem[{{Marsh} \& {Pop}(2015)}]{marsh15}
{Marsh}, D.~J.~E., \& {Pop}, A.-R. 2015, \mnras, 451, 2479

\bibitem[{{Martin} {et~al.}(2007){Martin}, {Ibata}, {Chapman}, {Irwin}, \&
  {Lewis}}]{martin07}
{Martin}, N.~F., {Ibata}, R.~A., {Chapman}, S.~C., {Irwin}, M., \& {Lewis},
  G.~F. 2007, \mnras, 380, 281

\bibitem[{{Mateo} {et~al.}(1993){Mateo}, {Olszewski}, {Pryor}, {Welch}, \&
  {Fischer}}]{mateo93}
{Mateo}, M., {Olszewski}, E.~W., {Pryor}, C., {Welch}, D.~L., \& {Fischer}, P.
  1993, \aj, 105, 510

\bibitem[{{Mateo} {et~al.}(2008){Mateo}, {Olszewski}, \& {Walker}}]{mateo08}
{Mateo}, M., {Olszewski}, E.~W., \& {Walker}, M.~G. 2008, \apj, 675, 201

\bibitem[{{McConnachie}(2012)}]{mcconnachie12}
{McConnachie}, A.~W. 2012, \aj, 144, 4

\bibitem[{{McConnachie} \& {C{\^o}t{\'e}}(2010)}]{mcconnachie10}
{McConnachie}, A.~W., \& {C{\^o}t{\'e}}, P. 2010, \apjl, 722, L209

\bibitem[{{McGaugh} \& {Milgrom}(2013)}]{mcgaugh13}
{McGaugh}, S., \& {Milgrom}, M. 2013, \apj, 766, 22

\bibitem[{{McGaugh}(2016)}]{mcgaugh16b}
{McGaugh}, S.~S. 2016, ArXiv:1610.06189

\bibitem[{{McGaugh} {et~al.}(2007){McGaugh}, {de Blok}, {Schombert}, {Kuzio de
  Naray}, \& {Kim}}]{mcgaugh07}
{McGaugh}, S.~S., {de Blok}, W.~J.~G., {Schombert}, J.~M., {Kuzio de Naray},
  R., \& {Kim}, J.~H. 2007, \apj, 659, 149

\bibitem[{{Milgrom}(1983)}]{milgrom83}
{Milgrom}, M. 1983, \apj, 270, 365

\bibitem[{{Olszewski} {et~al.}(1996){Olszewski}, {Pryor}, \&
  {Armandroff}}]{edo96}
{Olszewski}, E.~W., {Pryor}, C., \& {Armandroff}, T.~E. 1996, \aj, 111, 750

\bibitem[{{Plummer}(1911)}]{plummer11}
{Plummer}, H.~C. 1911, \mnras, 71, 460

\bibitem[{{Pontzen} \& {Governato}(2014)}]{pontzen14}
{Pontzen}, A., \& {Governato}, F. 2014, \nat, 506, 171

\bibitem[{{Read} {et~al.}(2016){Read}, {Agertz}, \& {Collins}}]{read16}
{Read}, J.~I., {Agertz}, O., \& {Collins}, M.~L.~M. 2016, \mnras, 459, 2573

\bibitem[{{Salucci} {et~al.}(2012){Salucci}, {Wilkinson}, {Walker}, {Gilmore},
  {Grebel}, {Koch}, {Frigerio Martins}, \& {Wyse}}]{salucci12}
{Salucci}, P., {Wilkinson}, M.~I., {Walker}, M.~G., {Gilmore}, G.~F., {Grebel},
  E.~K., {Koch}, A., {Frigerio Martins}, C., \& {Wyse}, R.~F.~G. 2012, \mnras,
  420, 2034

\bibitem[{{Sch{\"o}nrich} {et~al.}(2010){Sch{\"o}nrich}, {Binney}, \&
  {Dehnen}}]{schonrich10}
{Sch{\"o}nrich}, R., {Binney}, J., \& {Dehnen}, W. 2010, \mnras, 403, 1829

\bibitem[{{Shanks} {et~al.}(2015){Shanks}, {Metcalfe}, {Chehade}, {Findlay},
  {Irwin}, {Gonzalez-Solares}, {Lewis}, {Yoldas}, {Mann}, {Read}, {Sutorius},
  \& {Voutsinas}}]{shanks15}
{Shanks}, T., {Metcalfe}, N., {Chehade}, B., {Findlay}, J.~R., {Irwin}, M.~J.,
  {Gonzalez-Solares}, E., {Lewis}, J.~R., {Yoldas}, A.~K., {Mann}, R.~G.,
  {Read}, M.~A., {Sutorius}, E.~T.~W., \& {Voutsinas}, S. 2015, \mnras, 451,
  4238

\bibitem[{{Simon} \& {Geha}(2007)}]{simon07}
{Simon}, J.~D., \& {Geha}, M. 2007, \apj, 670, 313

\bibitem[{{Simon} {et~al.}(2011){Simon}, {Geha}, {Minor}, {Martinez}, {Kirby},
  {Bullock}, {Kaplinghat}, {Strigari}, {Willman}, {Choi}, {Tollerud}, \&
  {Wolf}}]{simon11}
{Simon}, J.~D., {Geha}, M., {Minor}, Q.~E., {Martinez}, G.~D., {Kirby}, E.~N.,
  {Bullock}, J.~S., {Kaplinghat}, M., {Strigari}, L.~E., {Willman}, B., {Choi},
  P.~I., {Tollerud}, E.~J., \& {Wolf}, J. 2011, \apj, 733, 46

\bibitem[{{Simon} {et~al.}(2016){Simon}, {Li}, {Drlica-Wagner}, {Bechtol},
  {Marshall}, {James}, {Wang}, {Strigari}, {Balbinot}, {Kuehn}, {Walker},
  {Abbott}, {Allam}, {Annis}, {Benoit-Levy}, {Brooks}, {Buckley-Geer}, {Burke},
  {Carnero Rosell}, {Carrasco Kind}, {Carretero}, {Cunha}, {D'Andrea}, {da
  Costa}, {DePoy}, {Desai}, {Doel}, {Fernandez}, {Flaugher}, {Frieman},
  {Garcia-Bellido}, {Gaztanaga}, {Goldstein}, {Gruen}, {Gutierrez},
  {Kuropatkin}, {Maia}, {Martini}, {Menanteau}, {Miller}, {Miquel}, {Neilsen},
  {Nord}, {Ogando}, {Plazas}, {Romer}, {Rykoff}, {Sanchez}, {Santiago},
  {Scarpine}, {Schubnell}, {Sevilla-Noarbe}, {Smith}, {Sobreira}, {Suchyta},
  {Swanson}, {Tarle}, {Whiteway}, \& {Yanny}}]{simon16}
{Simon}, J.~D., {Li}, T.~S., {Drlica-Wagner}, A., {Bechtol}, K., {Marshall},
  J.~L., {James}, D.~J., {Wang}, M.~Y., {Strigari}, L., {Balbinot}, E.,
  {Kuehn}, K., {Walker}, A.~R., {Abbott}, T.~M.~C., {Allam}, S., {Annis}, J.,
  {Benoit-Levy}, A., {Brooks}, D., {Buckley-Geer}, E., {Burke}, D.~L., {Carnero
  Rosell}, A., {Carrasco Kind}, M., {Carretero}, J., {Cunha}, C.~E.,
  {D'Andrea}, C.~B., {da Costa}, L.~N., {DePoy}, D.~L., {Desai}, S., {Doel},
  P., {Fernandez}, E., {Flaugher}, B., {Frieman}, J., {Garcia-Bellido}, J.,
  {Gaztanaga}, E., {Goldstein}, D.~A., {Gruen}, D., {Gutierrez}, G.,
  {Kuropatkin}, N., {Maia}, M.~A.~G., {Martini}, P., {Menanteau}, F., {Miller},
  C.~J., {Miquel}, R., {Neilsen}, E., {Nord}, B., {Ogando}, R., {Plazas},
  A.~A., {Romer}, A.~K., {Rykoff}, E.~S., {Sanchez}, E., {Santiago}, B.,
  {Scarpine}, V., {Schubnell}, M., {Sevilla-Noarbe}, I., {Smith}, R.~C.,
  {Sobreira}, F., {Suchyta}, E., {Swanson}, M.~E.~C., {Tarle}, G., {Whiteway},
  L., \& {Yanny}, B. 2016, ArXiv:1610.05301

\bibitem[{{Spergel} \& {Steinhardt}(2000)}]{spergel00}
{Spergel}, D.~N., \& {Steinhardt}, P.~J. 2000, Physical Review Letters, 84,
  3760

\bibitem[{{Strigari} {et~al.}(2008){Strigari}, {Bullock}, {Kaplinghat},
  {Simon}, {Geha}, {Willman}, \& {Walker}}]{strigari08}
{Strigari}, L.~E., {Bullock}, J.~S., {Kaplinghat}, M., {Simon}, J.~D., {Geha},
  M., {Willman}, B., \& {Walker}, M.~G. 2008, \nat, 454, 1096

\bibitem[{{Szentgyorgyi} {et~al.}(2011){Szentgyorgyi}, {Furesz}, {Cheimets},
  {Conroy}, {Eng}, {Fabricant}, {Fata}, {Gauron}, {Geary}, {McLeod}, {Zajac},
  {Amato}, {Bergner}, {Caldwell}, {Dupree}, {Goddard}, {Johnston}, {Meibom},
  {Mink}, {Pieri}, {Roll}, {Tokarz}, {Wyatt}, {Epps}, {Hartmann}, \&
  {Meszaros}}]{szentgyorgyi11}
{Szentgyorgyi}, A., {Furesz}, G., {Cheimets}, P., {Conroy}, M., {Eng}, R.,
  {Fabricant}, D., {Fata}, R., {Gauron}, T., {Geary}, J., {McLeod}, B.,
  {Zajac}, J., {Amato}, S., {Bergner}, H., {Caldwell}, N., {Dupree}, A.,
  {Goddard}, R., {Johnston}, E., {Meibom}, S., {Mink}, D., {Pieri}, M., {Roll},
  J., {Tokarz}, S., {Wyatt}, W., {Epps}, H., {Hartmann}, L., \& {Meszaros}, S.
  2011, \pasp, 123, 1188

\bibitem[{{The DES Collaboration} {et~al.}(2015){The DES Collaboration},
  {Drlica-Wagner}, {Bechtol}, {Rykoff}, {Luque}, {Queiroz}, {Mao}, {Wechsler},
  {Simon}, {Santiago}, {Yanny}, {Balbinot}, {Dodelson}, {Fausti Neto}, {James},
  {Li}, {Maia}, {Marshall}, {Pieres}, {Stringer}, {Walker}, {Abbott},
  {Abdalla}, {Allam}, {Benoit-Levy}, {Bernstein}, {Bertin}, {Brooks},
  {Buckley-Geer}, {Burke}, {Carnero Rosell}, {Carrasco Kind}, {Carretero},
  {Crocce}, {da Costa}, {Desai}, {Diehl}, {Dietrich}, {Doel}, {Eifler},
  {Evrard}, {Finley}, {Fosalba}, {Frieman}, {Gaztanaga}, {Gerdes}, {Gruen},
  {Gruendl}, {Gutierrez}, {Honscheid}, {Kuehn}, {Kuropatkin}, {Lahav},
  {Martini}, {Miquel}, {Nord}, {Ogando}, {Plazas}, {Reil}, {Roodman}, {Sako},
  {Sanchez}, {Scarpine}, {Schubnell}, {Sevilla-Noarbe}, {Smith},
  {Soares-Santos}, {Sobreira}, {Suchyta}, {Swanson}, {Tarle}, {Tucker},
  {Vikram}, {Wester}, {Zhang}, \& {Zuntz}}]{des15b}
{The DES Collaboration}, {Drlica-Wagner}, A., {Bechtol}, K., {Rykoff}, E.~S.,
  {Luque}, E., {Queiroz}, A., {Mao}, Y.-Y., {Wechsler}, R.~H., {Simon}, J.~D.,
  {Santiago}, B., {Yanny}, B., {Balbinot}, E., {Dodelson}, S., {Fausti Neto},
  A., {James}, D.~J., {Li}, T.~S., {Maia}, M.~A.~G., {Marshall}, J.~L.,
  {Pieres}, A., {Stringer}, K., {Walker}, A.~R., {Abbott}, T.~M.~C., {Abdalla},
  F.~B., {Allam}, S., {Benoit-Levy}, A., {Bernstein}, G.~M., {Bertin}, E.,
  {Brooks}, D., {Buckley-Geer}, E., {Burke}, D.~L., {Carnero Rosell}, A.,
  {Carrasco Kind}, M., {Carretero}, J., {Crocce}, M., {da Costa}, L.~N.,
  {Desai}, S., {Diehl}, H.~T., {Dietrich}, J.~P., {Doel}, P., {Eifler}, T.~F.,
  {Evrard}, A.~E., {Finley}, D.~A., {Fosalba}, P., {Frieman}, J., {Gaztanaga},
  E., {Gerdes}, D.~W., {Gruen}, D., {Gruendl}, R.~A., {Gutierrez}, G.,
  {Honscheid}, K., {Kuehn}, K., {Kuropatkin}, N., {Lahav}, O., {Martini}, P.,
  {Miquel}, R., {Nord}, B., {Ogando}, R., {Plazas}, A.~A., {Reil}, K.,
  {Roodman}, A., {Sako}, M., {Sanchez}, E., {Scarpine}, V., {Schubnell}, M.,
  {Sevilla-Noarbe}, I., {Smith}, R.~C., {Soares-Santos}, M., {Sobreira}, F.,
  {Suchyta}, E., {Swanson}, M.~E.~C., {Tarle}, G., {Tucker}, D., {Vikram}, V.,
  {Wester}, W., {Zhang}, Y., \& {Zuntz}, J. 2015, ArXiv:1508.03622

\bibitem[{{Torrealba} {et~al.}(2016){Torrealba}, {Koposov}, {Belokurov}, \&
  {Irwin}}]{torrealba16}
{Torrealba}, G., {Koposov}, S.~E., {Belokurov}, V., \& {Irwin}, M. 2016,
  \mnras, 459, 2370

\bibitem[{{Walker} {et~al.}(2009{\natexlab{a}}){Walker}, {Belokurov}, {Evans},
  {Irwin}, {Mateo}, {Olszewski}, \& {Gilmore}}]{walker09c}
{Walker}, M.~G., {Belokurov}, V., {Evans}, N.~W., {Irwin}, M.~J., {Mateo}, M.,
  {Olszewski}, E.~W., \& {Gilmore}, G. 2009{\natexlab{a}}, \apjl, 694, L144

\bibitem[{{Walker} \& {Loeb}(2014)}]{walker14}
{Walker}, M.~G., \& {Loeb}, A. 2014, Contemporary Physics, 55, 198

\bibitem[{{Walker} {et~al.}(2008){Walker}, {Mateo}, \& {Olszewski}}]{walker08}
{Walker}, M.~G., {Mateo}, M., \& {Olszewski}, E.~W. 2008, \apjl, 688, L75

\bibitem[{{Walker} {et~al.}(2007){Walker}, {Mateo}, {Olszewski}, {Gnedin},
  {Wang}, {Sen}, \& {Woodroofe}}]{walker07b}
{Walker}, M.~G., {Mateo}, M., {Olszewski}, E.~W., {Gnedin}, O.~Y., {Wang}, X.,
  {Sen}, B., \& {Woodroofe}, M. 2007, \apjl, 667, L53

\bibitem[{{Walker} {et~al.}(2009{\natexlab{b}}){Walker}, {Mateo}, {Olszewski},
  {Pe{\~n}arrubia}, {Wyn Evans}, \& {Gilmore}}]{walker09d}
{Walker}, M.~G., {Mateo}, M., {Olszewski}, E.~W., {Pe{\~n}arrubia}, J., {Wyn
  Evans}, N., \& {Gilmore}, G. 2009{\natexlab{b}}, \apj, 704, 1274

\bibitem[{{Walker} {et~al.}(2009{\natexlab{c}}){Walker}, {Mateo}, {Olszewski},
  {Sen}, \& {Woodroofe}}]{walker09b}
{Walker}, M.~G., {Mateo}, M., {Olszewski}, E.~W., {Sen}, B., \& {Woodroofe}, M.
  2009{\natexlab{c}}, \aj, 137, 3109

\bibitem[{{Walker} {et~al.}(2010){Walker}, {McGaugh}, {Mateo}, {Olszewski}, \&
  {Kuzio de Naray}}]{walker10}
{Walker}, M.~G., {McGaugh}, S.~S., {Mateo}, M., {Olszewski}, E.~W., \& {Kuzio
  de Naray}, R. 2010, \apjl, 717, L87

\bibitem[{{Walker} {et~al.}(2015){Walker}, {Olszewski}, \& {Mateo}}]{walker15}
{Walker}, M.~G., {Olszewski}, E.~W., \& {Mateo}, M. 2015, \mnras, 448, 2717

\bibitem[{{Walker, Mateo \& Olszewski}(2009)}]{walker09a}
{Walker, Mateo \& Olszewski}. 2009, \aj, 137, 3100

\bibitem[{{Wetzel} {et~al.}(2016){Wetzel}, {Hopkins}, {Kim},
  {Faucher-Gigu{\`e}re}, {Kere{\v s}}, \& {Quataert}}]{wetzel16}
{Wetzel}, A.~R., {Hopkins}, P.~F., {Kim}, J.-h., {Faucher-Gigu{\`e}re}, C.-A.,
  {Kere{\v s}}, D., \& {Quataert}, E. 2016, \apjl, 827, L23

\bibitem[{{Willman} \& {Strader}(2012)}]{willman12}
{Willman}, B., \& {Strader}, J. 2012, \aj, 144, 76

\bibitem[{{Wolf} {et~al.}(2010){Wolf}, {Martinez}, {Bullock}, {Kaplinghat},
  {Geha}, {Mu{\~n}oz}, {Simon}, \& {Avedo}}]{wolf10}
{Wolf}, J., {Martinez}, G.~D., {Bullock}, J.~S., {Kaplinghat}, M., {Geha}, M.,
  {Mu{\~n}oz}, R.~R., {Simon}, J.~D., \& {Avedo}, F.~F. 2010, \mnras, 406, 1220

\bibitem[{{Zhao}(1996)}]{zhao96}
{Zhao}, H. 1996, \mnras, 278, 488

\end{thebibliography}

\end{document}